\documentstyle{article}
\newcommand{\be}{\begin{equation}}
\newcommand{\ee}{\end{equation}}

\title{PREPOTENTIALS AND RIEMANN SURFACES}
\author{Robert Carroll\\University of Illinois\\ email:  
rcarroll@math.uiuc.edu}

\date{February, 1998}

\begin{document}
\bibliographystyle{plain}
\maketitle
\begin{abstract} 
We organize and review some material from various sources
about prepotentials, Riemann surfaces and 
kernels, WDVV, and the renormalization group, provide some further
connections and information, and indicate some directions and problems.
\end{abstract}

\section{INTRODUCTION}
\renewcommand{\theequation}{1.\arabic{equation}}\setcounter{equation}{0}

We extract here extensively from the important paper \cite{gf} where a 
profusion of relations between differentials on 
certain Riemann surfaces (RS)
 and the prepotential for Seiberg-Witten (SW) theory are exhibited.
This leads one to a complete determination of the prepotential (and
associated SW differential) via the structure of the RS.  A crucial
ingredient here is the generating functional of differentials
\be
W(\xi,\zeta)=\sum_1^{\infty}n\zeta^{n-1}d\zeta d\Omega_n(\xi)=
\partial_{\xi}\partial_{\zeta}log\,E(\xi,\zeta)
\label{71}
\ee
(where $E$ is the Fay prime form)
and its use is reminiscent of the complete determination of the free
energy $F$ for dKP via a certain kernel
\be
K(\mu,\lambda)=\frac{1}{P(\mu)-P(\lambda)}=\sum_1^{\infty}\partial_P
Q_n\mu^{-n};\,\,nQ_n={\cal B}_n=\lambda^n_{+}=\partial_nS
\label{C}
\ee
in \cite{ch,cf} (cf. also \cite{Aa,ci,cj,gh,gb,kw,kx,ta}).  We showed in 
\cite{cf} that $K$ has an analogue on a RS which can be written in terms
of a local Cauchy kernel based on the prime form $E$ and its construction
involves the generating functional $W$.  There is a slight difference
in notation in \cite{cf} due to using $d\Omega_n(\xi)\sim 
(-n\xi^{-n-1}-\sum_1^{\infty}q_{mn}\xi^{m-1})d\xi$, while in 
\cite{gf} $d\Omega_n=(\xi^{-n-1}+O(1))d\xi$ is used (thus a factor of
$(-n)$ arises in $W$ of (\ref{71}) which is not present in \cite{cf}).
Both $W$ and the RS analogue of $K$ are based on a one 
puncture situation so
the connection is legitimate and meaningful; the development for the two
puncture SW situation is indicated below for $W$ following \cite{gf}.
Connections to $K$ will also be exhibited in Section 6.
We also indicate relations between \cite{gf} and 
\cite{cb,ka,kc,ko,na} leading
to a discussion of WDVV for the Whitham theory and its relation to WDVV
for SW theory as in \cite{By,mr,Mw,Ms,Mz}.  Some relations between
\cite{Ms,Mz} and the Egorov geometry of \cite{dc} are also indicated.
We do not deal with prepotentials and master equations in the more
general context of string compactifications and Calabi-Yau (CY) manifolds
(see e.g. \cite{ky}).

\section{BACKGROUND FROM \cite{gf}}
\renewcommand{\theequation}{2.\arabic{equation}}\setcounter{equation}{0}

We take a SW situation following \cite{By,ba,Bz,cc,cg,cb,dg,de,da,ea,
gc,ia,kz,ko,la,me,mg,ma,na,sd,se} and recall the SW curves $\Sigma_g\,\,
(g=N-1)$ for a pure $SU(N)$ susy YM theory
\be
det_{N\times N}[L(w)-\lambda]=0;\,\,
P(\lambda)=\Lambda^N\left(w+
\frac{1}{w}\right);
\label{58}
\ee
$$
P(\lambda)=\lambda^N-\sum_2^Nu_k\lambda^{N-k}=\prod_1^N(\lambda-\lambda_j);
\,\,u_k=(-1)^k\sum_{i_1<\cdots<i_k}\lambda_{i_1}\cdots\lambda_{i_k}$$
Here the $u_k$ are Schur polynomials of $h_k=(1/k)\sum_1^N\lambda_i^k$ via
the formula
${\bf (A)}\,\,log(\lambda^{-N}P(\lambda))=-\sum_k(h_k/\lambda^k)$;
there are $g=N-1$ moduli $u_k$ and we refer to \cite{ia} for the Lax
operator $L$.
Thus $u_0=1,\,\,u_1=0,\,\,u_2=h_2,\,\,u_3=h_3,\,\,u_4=
h_4-(1/2)h_2^2,$ etc. ($h_1=0$ for $SU(N)$).
One also has the representation
\be
y^2=P^2(\lambda)-4\Lambda^{2N};\,\,y=\Lambda^N\left(w-\frac{1}{w}\right)
\label{60}
\ee
giving a two fold covering of the punctured Riemann sphere with parameter
$\lambda$.  Such Toda chain curves are characterized by a function
${\bf (B)}\,\,2\Lambda^Nw=(P+y)$.
Note also from (\ref{58}) - (\ref{60}) one obtains 
\be
\delta P+P'\delta\lambda=NP\delta log(\Lambda)+y\frac{\delta w}{w};\,\,
\delta P=-\sum\lambda^{N-k}\delta u_k;\,\,P'=\frac{\partial P}
{\partial\lambda}
\label{61}
\ee 
On a given curve (fixed $u_k$ and $\lambda$)
\be
\frac{dw}{w}=\frac{P'd\lambda}{y};\,\,dS_{SW}=\lambda\frac{dw}{w}=
\frac{\lambda dP}{y}
\label{62}
\ee
\be
\left.\frac{\partial dS_{SW}}{\partial u_k}\right|_{w=c}=\frac
{\lambda^{N-k}}{P'}\frac{dw}{w}=\frac{\lambda^{N-k}d\lambda}{y}=dv^k;
\,\,k=2,\cdots, N
\label{63}
\ee
where the $dv^k$ are $g=N-1$ holomorphic one forms with 
\be
a_i=\oint_{A_i}dS_{SW};\,\,
\sigma^{ik}=\oint_{A_i}dv^k=\frac{\partial a_i}{\partial u_k}
\label{64}
\ee
and $d\omega_i=(\sigma_{ik})^{-1}dv^k$ are the canonical holomorphic
differentials with ${\bf (C)}\,\,
\oint_{A_i}d\omega_j=\delta_{ij};\,\,\oint_{B_i}d\omega_j=B_{ij}$.
Note in (\ref{63}) it is necessary to assume $w$ is constant when the
moduli $u_k$ are varied in order to have $\partial 
dS_{SW}/\partial u_k=dv^k$ 
holomorphic.  
Now the periods $a_i=\oint_{A_i}dS_{SW}$ define the $a_i$ as 
functions of $u_k$ (i.e. $h_k$) and $\Lambda$, or inversely, the $u_k$
as functions of $a_i$ and $\Lambda$.
One proves for example (see below)
\be
\frac{\partial u_k}{\partial\,log(\Lambda)}=
ku_k-a_i\frac{\partial u_k}{\partial 
a_i}
\label{70}
\ee
and generally $\Lambda$ and $T_1$ can be identified after suitable
scaling (cf. \cite{By,ba,Bz,cb,ea,gf}).  This is in keeping with the
idea in \cite{ea} that Whitham times are used to restore the homogeneity
of the prepotential when it is disturbed by renormalization (cf. also
\cite{cb}).
\\[3mm]\indent
For the prepotential
one goes to \cite{ia,na} for example and
defines differentials ${\bf (D)}\,\,
d\Omega_n\sim (\xi^{-n-1}+O(1))d\xi$ for $n\geq 1$ with
$\oint_{A_i}d\Omega_n=0$ (pick one puncture momentarily).  
This leads to $W$ in (\ref{71}) where
\be
W(\xi,\zeta)\sim\frac{d\xi d\zeta}{(\xi-\zeta)^2}+O(1)=\sum_1^{\infty}
n\frac{d\xi}{\xi^{n+1}}\zeta^{n-1}d\zeta +O(1)
\label{72}
\ee
One can also
impose a condition of the form
${\bf (E)}\,\,\partial d\hat{\Omega}/\partial\,(moduli)=
holomorphic$ on differentials $d\hat{\Omega}_n$
and use a generating functional (note we distinguish $dS$ and
$dS_{SW}$)
\be
dS=\sum_1^{\infty}T_nd\hat{\Omega}_n=\sum_1^g\alpha_id\omega_i+\sum_0^
{\infty}T_nd\Omega_n
\label{73}
\ee
(we have added a $T_0d\Omega_0$ term here even though $T_0=0$ in the
pure $SU(N)$ theory - cf. \cite{ea} and Section 3).  Note that there
is a possible confusion in notation with $d\Omega_n$ since we will
choose the $d\hat{\Omega}_n$ below to have poles at $\infty_{\pm}$
and the poles must balance in (\ref{73}).  The matter is 
clarified by noting
that {\bf (D)} determines singularities at $\xi=0$ and for our curve
$\xi=0\sim \infty_{\pm}$ via $\xi=w^{\mp 1/N}$ which in turn corresponds
to $P(\lambda)^{1/N}$ for $\Lambda =1$
(see (\ref{83}), (\ref{95}), and remarks before
(\ref{80}) - cf. also (\ref{96}) for $d\Omega_n^{\pm}$).  
Thus in a certain
sense $d\Omega_n$ here must correspond to $d\Omega_n^{+}+d\Omega_n^{-}$ 
in a hyperelliptic parametrization (cf. (\ref{96})) and
$T_n\sim T_n^{+}=T_n^{-}$ (after adjustment for the 
singular coefficient at
$\infty_{\pm}$).
This would also be consistent with symmetrization in (\ref{10}) (cf.
Summary 3.2).
The periods $\alpha_i=\oint_{A_i}dS$ can be considered as coordinates
on the moduli space (note these are not the $\alpha_i$ of \cite{cc,na}).  
They are not just the same as the $a_i$ but are
defined as functions of $h_k$ and $T_n$ (or alternatively $h_k$ can
be defined as functions of $\alpha_i$ and $T_n$ so that derivatives
$\partial h_k/\partial T_n$ for example are nontrivial.  One will consider
the variables $\alpha_i$ and $T_n=Res_{\xi=0}\xi^ndS(\xi)$ as independent
so that 
\be
\frac{\partial dS}{\partial \alpha_i}=d\omega_i;\,\,\frac{\partial dS}
{\partial T_n}=d\Omega_n
\label{74}
\ee
Next one can introduce the prepotential $F(\alpha_i,T_n)$ via an analogue
of $a_i^D=\partial{\cal F}/\partial a_i$, namely
\be
\frac{\partial F}{\partial \alpha_i}=\oint_{B_i}dS;\,\,\frac{\partial F}
{\partial T_n}=\frac{1}{2\pi in}Res_0\xi^{-n}dS
\label{75}
\ee
Then one notes the formulas
\be
\frac{\partial^2 F}{\partial T_m\partial T_n}=\frac{1}{2\pi in}
Res_0\xi^{-n}\frac{\partial dS}{\partial T_m}=\frac{1}{2\pi in}
Res_0\xi^{-n}d\Omega_m=\frac{1}{2\pi im}Res_0\xi^{-m}d\Omega_n
\label{76}
\ee
(the choice of $\xi$ is restricted to $w^{\pm 1/N}$ in the situation of
(\ref{60}))
and factors
like $n^{-1}$ arise since $\xi^{-n-1}d\xi=-d(\xi^{-n}/n)$.  Below
we use also a slightly different normalization $d\Omega_n\sim
\pm w^{\pm n/N}(dw/w)=(N/n)dw^{\pm n/N}$ near $\infty_{\pm}$
so that residues in (\ref{75}) and (\ref{76})
will be multiplied by $N/n$ instead of $1/n$.  
Note also that in accord with the remarks above
$Res_{\xi=0}$ will correspond to the sum of residues at 
$\infty_{\pm}$ involving $\xi=w^{\mp 1/N}$ which in turn corresponds to
$P(\lambda)^{1/N}$. 
By definition
$F$ is a homogeneous function of $\alpha_i$ and $T_n$ of degree two, so that
\be
2F=\alpha_i\frac{\partial F}{\partial \alpha_i}+T_n\frac{\partial F}
{\partial T_n}=\alpha_i\alpha_j\frac{\partial^2F}{\partial\alpha_i\partial
\alpha_j}+2\alpha_iT_n\frac{\partial^2F}{\partial\alpha_i\partial T_n}
+T_nT_m\frac{\partial^2F}{\partial T_n\partial T_m}
\label{78}
\ee
Note however 
that $F$ is not just a quadratic function of $\alpha_i$ and $T_n$; a
nontrivial dependence on these variables arises through the dependence
of $d\omega_i$ and $d\Omega_n$ on the moduli (such as $u_k$ or $h_k$)
which in turn depend on $\alpha_i$ and $T_n$.  The dependence is described
by a version of Whitham equations, obtained for example by substituting
(\ref{73}) into (\ref{74}).  Thus
\be
d\hat{\Omega}_n+T_m\frac{\partial d\hat{\Omega}_n}{\partial u_{\ell}}
\frac{\partial u_{\ell}}{\partial T_m}=d\Omega_n\Rightarrow
\left(\sum_{m,\ell}T_m\frac{\partial u_{\ell}}{\partial T_m}\right)
\oint_{A_i}\frac{\partial d\hat{\Omega}_n}{\partial u_{\ell}}=
-\oint_{A_i}d\hat{\Omega}_n
\label{79}
\ee
(since $\oint_{A_I}d\Omega_n=0$). 
By {\bf (E)} one has $\oint_{A_i}(\partial 
dS/\partial u_{\ell})=\oint_{A_I}dV_i=\Sigma_{i\ell}$ where $dV_{\ell}\sim$
holomorphic, and hence from (\ref{73}) there results ${\bf (F)}\,\,
T_m(\partial u_k/\partial T_m)=\Sigma_{ki}^{-1}\alpha_i$, which furnishes
the Whitham dynamics for $u_k$ (cf. also \cite{ia}).
\\[3mm]\indent
Now the SW spectral curves (\ref{60}) are related to Toda hierarchies
with two punctures.  We recall (\ref{58}) - (\ref{60}) and note
therefrom that ${\bf (G)}\,\,w^{\pm 1}=(1/2\Lambda^N)(P\pm y)
\sim (1/\Lambda^N)
P(\lambda)(1+O(\lambda^{-2N}))$ near $\lambda=\infty_{\pm}$ since
from $y^2=P^2-4\Lambda^{2N}$ we have $(y/P)=[1-(4\Lambda^{2N}/P^2)]^{1/2}
=(1+O(\lambda^{-2N}))$ so $w^{\pm 1}=(P/2\Lambda^N)[1\pm (y/P)]=
(P/2\Lambda^N)[2+O(\lambda^{-2N})]$.
One writes
$w(\lambda=\infty_{+})=\infty$ and $w(\lambda=\infty_{-})=0$ with
$\xi\sim w^{\mp1/N}$ (i.e. $\xi=w^{-1/N}\sim\lambda^{-1}$ at 
$\infty_{+}$ and
$\xi=w^{1/N}\sim \lambda^{-1}$ at $\infty_{-}$).  
Near $\lambda=\pm\infty$ by {\bf (G)} one can write then $w^{\pm 1/N}\sim
P(\lambda)^{1/N}$ (for $\Lambda=1$)
in calculations involving $w^{\pm n/N}$ with $n<2N$.
The $w$ parametrization
is of course not hyperelliptic and we note that {\bf (D)} applies for
$\xi=w^{\pm 1/N}$ with all $d\Omega_n$ and later only for 
$\xi=\lambda^{-1}$ 
in certain differentials $d\tilde{\Omega}_n$.
It would now be possible to envision
differentials
$d\Omega_n^{\pm}$ and $d\hat{\Omega}_n^{\pm}$ but
$d\hat{\Omega}_n=d\hat{\Omega}_n^{+}+d\hat{\Omega}_n^{-}$ is then clearly
the only admissible object (i.e. $d\hat{\Omega}_n$ must have poles at
both punctures); this is suggested
by the form $dw/w$ in (\ref{62}) and  
the
coefficients of $w^{n/N}$ at $\infty_{+}$ and of $w^{-n/N}$ at $\infty_{-}$
must be equal (see also remarks above about $d\Omega_n$ in (\ref{73})).
This corresponds to the Toda chain situation with the same
dependence on plus and minus times.  Moreover 
one takes differentials $d\hat{\Omega}_n$ for (\ref{60})
($\Lambda=1$ here for awhile to simplify formulas)
\be
d\hat{\Omega}_n=R_n(\lambda)\frac{dw}{w}=P_{+}^{n/N}(\lambda)\frac{dw}{w}
\label{80}
\ee
These differentials satisfy {\bf (E)} provided the moduli derivatives
are taken at constant $w$ (not $\lambda$) and we can use the formalism
developed above for $\xi=w^{\mp 1/N}$.  The SW differential $dS_{SW}$ is
then simply $dS_{SW}=d\hat{\Omega}_1$, i.e.
\be
\left. dS\right|_{T_n=\delta_{n,1}}=dS_{SW};\,\,\left.\alpha_i\right|_{T_n=
\delta_{n,1}}=a_i;\,\,\left.\alpha_i^D\right|_{T_n=\delta_{n,1}}=a_i^D
\label{82}
\ee
With this preparation we can now write
for $n<2N$, with $d\hat{\Omega}_n$ as indicated
in (\ref{80})
\be
\frac{\partial F}{\partial T_n}=\frac{N}{2\pi in}\left(Res_{\infty_{+}}
w^{n/N}dS+Res_{\infty_{-}}w^{-n/N}dS\right)=
\label{83}
\ee
$$=\frac{N}{2\pi in}\left(Res_{\infty_{+}}w^{n/N}+
Res_{\infty_{-}}w^{-n/N}\right)\left(\sum_m
T_mP_{+}^{m/N}(\lambda)\right)\frac{dw}{w}=$$
$$=\frac{N^2}{i\pi n^2}\sum_mT_mRes_{\infty}\left(P_{+}^{m/N}(\lambda)dP^
{n/N}(\lambda)\right)=-\frac{N^2}{i\pi n^2}\sum_mT_mRes_{\infty}
\left(P^{n/N}(\lambda)dP_{+}^{m/N}(\lambda)\right)$$
One can introduce Hamiltonians here of great importance in the general
theory (cf. \cite{cb,Da,gf,ia,lg}) but we only indicate a few relations
since our present concerns lie elsewhere.  Then
evaluating at $T_n=\delta_{n,1}$ (\ref{83})
becomes
\be
\frac{\partial F}{\partial T_n}=-\frac{N^2}{i\pi n^2}Res_{\infty}
P^{n/N}(\lambda)d\lambda=\frac{N}{i\pi n}{\cal H}_{n+1}
\label{85}
\ee
where 
$${\cal H}_{n+1}=-\frac{N}{n}Res_{\infty}P^{n/N}d\lambda=\sum_{k\geq 1}
\frac{(-1)^{k-1}}{k!}\left(\frac{n}{N}\right)^{k-1}\sum_{i_1+
\cdots +i_k=n+1}
h_{i_1}\cdots h_{i_k}=$$
\be
=h_{n+1}-\frac{n}{2N}\sum_{i+j=n+1}h_ih_j +O(h^3)
\label{86}
\ee
This can be rephrased as
\be
\frac{\partial F}{\partial T_n}=\frac{\beta}{2\pi in}
\sum_mmT_m{\cal H}_{m+1,n+1}=
\frac{\beta}{2\pi in}T_1{\cal H}_{n+1}+O(T_2,T_3,\cdots)
\label{87}
\ee
where
\be
{\cal H}_{m+1,n+1}=-\frac{N}{mn}Res_{\infty}
\left(P^{n/N}dP_{+}^{m/N}\right)=
-{\cal H}_{n+1,m+1};
\label{91}
\ee
$${\cal H}_{n+1}\equiv {\cal H}_{n+1,2}=-\frac{N}{n}Res_{\infty}
P^{n/N}d\lambda=h_{n+1}+O(h^2)$$
\\[3mm]\indent
Now for the mixed derivatives one writes
\be
\frac{\partial^2F}{\partial \alpha_i\partial T_n}=\oint_{B_i}d\Omega_n=
\frac{1}{2\pi in}Res_0\xi^{-n}d\omega_i=
\label{88}
\ee
$$=\frac{N}{2\pi in}\left(Res_{\infty_{+}}w^{n/N}d\omega_i+
Res_{\infty_{-}}
w^{-n/N}d\omega_i\right)=\frac{N}{i\pi n}Res_{\infty}P^{n/N}d\omega_i$$
Next set ${\bf (H)}\,\,P^{n/N}=\sum_{-\infty}^{\infty}p^N_{nk}
\lambda^k$ so that ${\bf (I)}\,\,Res_{\infty}P^{n/N}d\omega_i=
\sum_{-\infty}^n
p^N_{nk}Res_{\infty}\lambda^kd\omega_i$.  Then e.g.
$$
d\omega_j(\lambda)=\sigma_{jk}^{-1}dv^k(\lambda)=
\sigma_{jk}^{-1}\frac{\lambda^{N-k}d\lambda}{y(\lambda)}=
\sigma_{jk}^{-1}\frac{\lambda^{N-k}d\lambda}{P(\lambda)}\left(1+O(\lambda^
{-2N})\right)=$$
\be
=-\sigma_{jk}^{-1}\frac{\partial\,log\,P(\lambda)}{\partial u_k}
d\lambda\left(1+O(\lambda^{-2N})\right)
\label{89}
\ee
From {\bf (A)} and $\sigma_{jk}^{-1}=\partial u_k/\partial a_j$ one
obtains then
\be
d\omega_j(\lambda)\left(1+O(\lambda^{-2N})\right)=\sum_{n\geq 2}
\sigma_{jk}^
{-1}\frac{\partial h_n}{\partial u_k}\frac{d\lambda}{\lambda^n}=
\sum_{n\geq 1}\frac{\partial h_{n+1}}{\partial a_i}\frac{d\lambda}
{\lambda^{n+1}}
\label{90}
\ee
so for $k <2N,\,\,{\bf (J)}\,\,Res_{\infty}\lambda^kd\omega_i=
\partial h_{k+1}/\partial a_i$.  
Further analysis yields an equation
${\bf (K)}\,\,Res_{\infty}w^{n/N}d\omega_i
\newline
=Res_{\infty}P^{n/N}
d\omega_i=\partial {\cal H}_{n+1}/\partial a_i$ leading to
\be
\frac{\partial^2F}{\partial \alpha_i\partial T_n}=\frac{N}{i\pi n}Res_
{\infty}P(\lambda)^{n/N}d\omega_i=\frac{N}{i\pi n}\frac{\partial
{\cal H}_{n+1}}{\partial \alpha_i}
\label{94}
\ee
For the second $T$ derivatives one uses the general 
formula (\ref{76}) written as
\be
\frac{\partial^2F}{\partial T_n\partial T_m}=\frac{1}{2\pi in}Res_0
\xi^{-n}d\Omega_m=\frac{N}{2\pi in}\left(Res_{\infty_{+}}w^{n/N}d\Omega_m
+Res_{\infty_{-}}w^{-n/N}d\Omega_m\right)
\label{95}
\ee
while for the second $\alpha$ derivatives one has evidently
\be
\frac{\partial^2F}{\partial\alpha_i\partial\alpha_j}=
\oint_{B_i}d\omega_j=B_{ij}
\label{300}
\ee
Note that one can also use differentials $d\tilde{\Omega}$ defined
by {\bf (D)} with $\xi=\lambda^{-1}$ 
(not $\xi=w^{\mp 1/N}$); recall $\infty_{\pm}\sim (\pm,\lambda\to\infty)$
in the hyperelliptic parametrization. 
This
leads to
\be
d\Omega_n^{\pm}\sim\pm\left(w^{\pm n/N}+O(1)\right)\frac{dw}{w}=
\frac{N}{n}dw^{\pm n/N}+\cdots =
\label{96}
\ee
$$=\frac{N}{n}dP^{n/N}+\cdots =\frac{N}{n}\sum_1^nkp^N_{nk}\lambda^{k-1}
d\lambda+\cdots =\frac{N}{n}\sum_1^nkp^N_{nk}d\tilde{\Omega}_k^{\pm}$$
Putting 
{\bf (H)} and (\ref{96}) into (\ref{95}) gives then
\be
\frac{\partial^2F}{\partial T_m\partial T_n}=-\frac{N^2}{i\pi mn}
\sum_1^m\ell p^N_{m\ell}Res_{\infty}w^{n/N}d\tilde{\Omega}_{\ell}
\label{97}
\ee
where $d\tilde{\Omega}_{\ell}=d\tilde{\Omega}_{\ell}^{+}+
d\tilde{\Omega}_{\ell}^{-}$.
\\[3mm]\indent
Further analysis in \cite{gf} involves theta functions and the Szeg\"o
kernel (cf. \cite{fa,gf}).  Thus let $E$ be the 
even theta characteristic associated
with the distinguished separation of ramification points into two equal
sets $P(\lambda)\pm 2\Lambda^N=\prod_1^N(\lambda-r_{\alpha}^{\pm})$.  This
allows one to write the square of the corresponding Szeg\"o kernel as
\be
\Psi_E^2(\lambda,\mu)=\frac{P(\lambda)P(\mu)-4\Lambda^{2N}+y(\lambda)
y(\mu)}{2y(\lambda)y(\mu)}\frac{d\lambda d\mu}{(\lambda-\mu)^2}
\label{99}
\ee
We can write (cf. \cite{gf})
\be
\Psi_E^2(\lambda,\mu)=\sum_{n\geq 1}\hat{\Psi}_E^2(\lambda)\frac{n\lambda^
{n-1}d\mu}{\mu^{n+1}}\left(1+O(P^{-1}(\mu)\right);
\label{102}
\ee
$$\hat{\Psi}_E^{\pm}(\lambda)\equiv\frac{P\pm y}{2y}d\lambda=\left\{
\begin{array}{cc}
(1+O(\lambda^{-2N}))d\lambda & near\,\,\infty_{\pm}\\
O(\lambda^{-2N}d\lambda & near\,\,\infty_{\mp}
\end{array}\right.$$
and
utilize the formula
\be
\Psi_E(\xi,\zeta)\Psi_{-E}(\xi,\zeta)=W(\xi,\zeta)+d\omega_i(\xi)d\omega_j
(\zeta)\frac{\partial^2}{\partial z_i\partial z_j}log\,\theta_E(\vec{0}|B)
\label{103}
\ee
(cf. \cite{fa,gf}).  Here one uses (\ref{90}) and (\ref{71}) to get
($1\leq n <2N,\,\,\zeta\sim 1/\mu$)
$$
d\omega_j(\mu)=\sum_{n\geq 1}\frac{nd\mu}{\mu^{n+1}}\left(\frac{1}{n}
\frac{\partial h_{n+1}}{\partial a_j}\right);\,\,d\tilde{\Omega}_n^{\pm}
(\lambda)=\lambda^{n-1}\hat{\Psi}_E^{2\pm}(\lambda)-
\rho_n^id\omega_i(\lambda);
$$
\be
d\tilde{\Omega}_n(\lambda)=\lambda^{n-1}(\hat{\Psi}_E^{2+}
(\lambda)+\hat{\Psi}_
E^{2-}(\lambda))-2\rho_n^id\omega_i(\lambda)
\label{104}
\ee
where ${\bf (L)}\,\,\rho_n^i=(1/n)(\partial h_{n+1}/\partial a_j)
\partial^2_{ij}log\,\theta_E(\vec{0}|B)$.  
To use this we note
\be
\sum_k\left(\left.\frac{\partial u_k}{\partial\,log(\Lambda)}\right|_
{a_i=c}\right)\oint_{A_i}\frac{\partial dS_{SW}}{\partial u_k}+\oint_{A_i}
\frac{\partial dS_{SW}}{\partial\,log(\Lambda)}=0
\label{67}
\ee
Then there results
\be
\sum_k\frac{\partial u_k}{\partial\,log(\Lambda)}\frac{\partial a_i}
{\partial u_k}=-\oint_{A_i}\frac{\partial dS_{SW}}{\partial\,log(\Lambda)}=
-N\oint_{A_i}\frac{P}{P'}\frac{dw}{w}=-N\oint_{A_i}\frac{Pd\lambda}{y}=
\label{68}
\ee
$$=-N\oint_{A_i}\frac{P+y}{y}d\lambda=-2N\Lambda^N\oint_{A_i}\frac
{wd\lambda}{y}=-2N\Lambda^N\oint_{A_i}wdv^N$$
Here one is taking $dS_{SW}=\lambda dw/w$ and using (\ref{61}) in the form
($\delta P=\delta w=0$)
$\delta dS_{SW}/\delta\,log(\Lambda)=
\newline
\delta\lambda(dw/w)/\delta\,log)\Lambda)=
(NP/P')(dw/w)$.  Then (\ref{62})
gives $NPd\lambda/y$ and the next step involves $\oint_{A_i}d\lambda=0$.  
Next {\bf (B)} is used along with (\ref{63}).  Note also from
${\bf (M)}\,\,\lambda dP=\lambda[N\lambda^{N-1}-\sum (N-k)u_k\lambda^{N-k-1}]
d\lambda=NPd\lambda+\sum ku_k\lambda^{N-k}d\lambda$ one obtains via
(\ref{62}), (\ref{63}), and (\ref{68}) (middle term)
\be
-\sum\frac{\partial u_k}{\partial\,log(\Lambda)}\frac{\partial a_i}{\partial
u_k}=N\oint_{A_i}\frac{Pd\lambda}{y}=
\label{69}
\ee
$$= \oint_{A_i}\left(\frac{\lambda dP}{y}-\sum ku_k
\lambda^{N-k}\frac{d\lambda}
{y}\right)=a_i-\sum ku_k\frac{\partial a_i}{\partial u_k}$$
which evidently implies (\ref{70}).  Now from (\ref{68}) there results
\be
-\frac{\partial u_k}{\partial \,log(\Lambda)}\frac
{\partial a_i}{\partial u_k}
=2N\Lambda^N\oint_{A_i}\frac{wd\lambda}{y}=N\oint_{A_i}\frac{P+y}{y}
d\lambda=
\label{105}
\ee
$$= 2N\oint_{A_i}\hat{\Psi}^2_E(\lambda)=2N\rho_1^i=2N\frac{\partial h_2}
{\partial a_j}\partial^2_{ij}log\,\theta_E(\vec{0}|B)$$
from which
\be
\frac{\partial u_k}{\partial\,log(\Lambda)}=-2N\frac{\partial u_k}{\partial
a_i}\frac{\partial u_2}{\partial a_j}\partial^2_{ij}log\,\theta_E
(\vec{0}|B)
\label{106}
\ee
Here one can replace $u_k$ by any function of $u_k$ alone such as
$h_k$ or ${\cal H}_{n+1}$ (note $u_2=h_2$).  
Note also (cf. \cite{cb,dz,ea,gf,he,ma,se}) that identifying
$\Lambda$ and $T_1$
(after
appropriate rescaling $h_k\to T_i^kh_k$ and ${\cal H}_k\to T_1^k{\cal H}_k$)
one has ($\beta=2N$)
\be
\frac{\partial F_{SW}}{\partial\,log(\Lambda)}=
\frac{\beta}{2\pi i}(T_1^2h_2)
\label{100}
\ee
(this equation for $\partial F/\partial\,log(\Lambda)$ also follows
directly from (\ref{85}) - (\ref{87}) when $T_n=0$ for $n\geq 2$ 
since ${\cal H}_2=h_2$).
\\[3mm]\indent
Finally consider (\ref{61}) in the form ${\bf (N)}\,\,
P'\delta\lambda -\sum_k\lambda^{N-k}\delta u_k=NP\delta\,log(\Lambda)$.
There results (cf. (\ref{68}))
\be
\delta a_i=\oint_{A_i}\delta\lambda\frac{dw}{w}=\sum_k\delta u_k
\oint_{A_i}\frac{\lambda^{N-k}}{P'}\frac{dw}{w}+N\delta\,log(\Lambda)
\oint_{A_i}\frac{P}{P'}\frac{dw}{w};
\label{200}
\ee
$$\sum_k\oint_{A_i}dv^k\left(\left.\frac{\partial u_k}{\partial\,
log(\Lambda)}\right|_{a=\hat{c}}\right)=-N\oint_{A_i}
\frac{P}{P'}\frac{dw}{w}
=-N\oint_{A_i}\frac{Pd\lambda}{y}$$
On the other hand for $\alpha_i=T_1a_i+O(T_2,T_3,\cdots)$
\be
\delta\alpha_i=\alpha_i\delta\,log(T_1)+T_1\oint_{A_i}\delta\lambda
\frac{dw}{w}+O(T_2,T_3,\cdots)
\label{201}
\ee
so for constant $\Lambda$ with $T_n=0$ for $n\geq 2$
(while $\alpha_i$ and $T_n$ are independent)
\be
\sum_k\oint_{A_i}dv^k\left(\left.\frac{\partial u_k}{\partial\,log(T_1)}
\right|_{\alpha=c}\right)=-\frac{\alpha_i}{T_1}=
-\oint_{A_i}\frac{\lambda dP}{y}
\label{202}
\ee
Since $\lambda dP=NPd\lambda+\sum_kku_k\lambda^{N-k}d\lambda$ it follows that
(cf. (\ref{70}))
\be
\left.\frac{\partial u_k}{\partial\,log(T_1)}\right|_{\alpha=c}=
\left.\frac{\partial u_k}{\partial\,log(\Lambda)}\right|_{a=\hat{c}}-ku_k=
-a_i\frac{\partial u_k}{\partial a_i}
\label{203}
\ee
(cf. (\ref{70}) - note the evaluation points are different and
$\alpha_i=T_1a_i+O(T_2,T_3,\cdots)$).  
This relation is true for any homogeneous algebraic
combination of the $u_k$ (e.g. for $h_k$ and ${\cal H}_k$).
We will return to the $log(\Lambda)$ derivatives later in connection with
WDVV.  For further relations involving Whitham theory
and $\Lambda$ derivatives see \cite{bc,By,bd,ba,Bz,be,cb,dz,dg,ea,gf,
ia,sb} and references there.
\\[3mm]\indent
Let us now look at this material from a different point of view.  The 
discovery of SW curves and their connection to physics goes back to
\cite{sd} and has further mathematical connections involving Calabi-Yau
(CY) and mirror manifolds, special geometry, etc., to string theory and
branes (cf. \cite{cd,cn,de,da,fg,ge,Ga,kz,la,lg,ld,wa}).  What seems
to result is that at various stages there appear
purely mathematical constructions describing various subsets of some general
still unknown physical theory.  That mathematics should play 
such a fundamental
role is not of course surprising in view of historical developments in
mathematical physics (which are still being refined and updated).  To
describe the roles of RS in general physical theories seems therefore not
at all unnatural.  This is especially so in view of the solution of the
famous Schottky problem in \cite{mt,sa} (cf. \cite{cz,tb} for sketches,
references, etc.).  Thus roughly a principally polarized Abelian variety
is the Jacobian of a RS if and only if it is the orbit of a suitable
KP flow.  Such a connection between RS and KP flows is also evident from
the construction of a unique BA function on a general RS as in 
\cite{aa,bh,cc,Db} for example, which determines a finite zone 
(quasiperiodic)
KP flow.  This really says that KP flows are mathematical objects basically
(which can be seen also from their derivation via Kac-Moody (KM) algebras,
the Hirota bilinear identity, etc.) but there are also classical physical
meanings in terms of water waves for example as well as many connections
to conformal field theory (CFT) and other quantum mechanical disciplines.
In the present situation one looks at a subclass of RS described by
(\ref{58}) - (\ref{60}).  Given the RS (with a fixed homology basis)
the holomorphic differentials $d\omega_i$ and the $a_i$ variables
are completely determined.  Similarly, with the various parametrizations
indicated, the differentials $d\Omega_n^{\pm},\,\,d\hat{\Omega}_n,\,\,
d\Omega_n$,
and $d\tilde{\Omega}_n$ are determined, alone with the Szeg\"o kernel,
and there will be various connections between the 
$d\Omega_n^{\pm}$ and $d\omega_i$ via Riemann relations etc. 
(cf. \cite{cc,cf,sc} and Section 6).
\\[3mm]\indent
{\bf CONCLUSION 2.1.}$\,\,$  Given the RS (\ref{58}) - (\ref{60})
one can determine $\partial F/\partial T_n$ from (\ref{83}) - (\ref{85}),
$\partial^2F/\partial\alpha_i\partial T_n$ from (\ref{94}),
$\partial^2F/\partial T_n\partial T_m$ from (\ref{95}), and 
$\partial^2F/\partial\alpha_i\partial\alpha_j$ from (\ref{300}) (also
for $T_n=\delta_{n,1},\,\,\alpha_i=a_i$ as in (\ref{82})
with $\Lambda =1$).  Finally
the equation (\ref{100}) for $\partial F/\partial\,log(\Lambda)$
corresponds to an identification $T_1\sim\Lambda$ and follows from
(\ref{85}) - (\ref{87}) when $T_n=0$ for $n\geq 2$.  Therefore,
since $F\sim F_{SW}$ for $T_1=1$ we see that all derivatives of 
$F_{SW}$ are determined by the RS alone so up to a normalization the 
prepotential is completely determined by the RS.  We emphasize
that $F_{SW}$ involves basically $T_n=\delta_{n,1}$ 
only, with no higher $T_n$, and
$\alpha_i=a_i$, whereas $F$ involving $\alpha_i$ and $T_n$ is defined
for all $T_n$ (cf. here \cite{gf}).
\\[3mm]\indent {\bf REMARK 2.2.}$\,\,$  In \cite{gf} one writes
\be
F_{GKM}^N(T)=\sum_{m,n}\frac{T_mT_n}{2mn}Res_{\infty}\left(P^{n/N}dP_{+}^
{m/N}(\lambda)\right);
\label{84}
\ee
$$T_n^{GKM}=-\frac{N}{N-n}Res_{\infty}P(\lambda)^{1-(n/N)}d\lambda$$
where GKM refers to an apparently associated generalized Kontsevich
model (cf. \cite{ke,kf,mc} - we omit all details about GKM matrix
models).  Some calculations based on (\ref{95}) and (\ref{97}) leads
then to
\be
F_{GKM}^N(\alpha,T)\equiv\frac{1}{2N}\sum_{m,n}T_mT_n{\cal H}_{m+1,n+1}
\label{302}
\ee
\be
\frac{\partial^2}{\partial T_m\partial T_n}\left(F(\alpha,T)-\frac
{\beta^2}{4\pi i}F_{GKM}^N(\alpha,T)\right)=-\frac{\beta^2}{2\pi inm}
\frac{\partial{\cal H}_{m+1}}{\partial a_i}\frac{\partial{\cal H}_{n+1}}
{\partial a_j}\partial^2_{ij}log\theta_E(\vec{0}|B)
\label{303}
\ee
There is an extensive theory of GKM in \cite{ke,kf,mc} for example
involving Miwa times and Whitham times but it is not entirely clear what
model is referred to here.  For LG models one might think of $P(\lambda)
\sim W'(\lambda)$ where $W$ is a superpotential.  The SW theory 
is apparently
a kind of 1-loop generalization of the Kontsevich theory.
\\[3mm]\indent {\bf REMARK 2.3.}$\,\,$  Going to \cite{By} we see that
in the present context one can define a beta function matrix (for
$h_k=(1/k)<Tr\phi^k>$ fixed)
\be
\beta_{ij}=\Lambda\frac{\partial\tau_{ij}}{\partial\Lambda};\,\,
\tau_{ij}=\frac{\partial^2F_{SW}}{\partial a_i\partial a_j}
\label{304}
\ee
where $i,j=1,\cdots,N-1$.  We will think here of $T_1\sim\Lambda$ so there
are no adjoined $T_n$ variables, 
only $a_i\,\,(1\leq i\leq N-1$) and $\Lambda$.
One introduces now a new variable $a_0$ and defines $\partial_0F_{SW}=
\partial_{\Lambda}F_{SW}$; in comparing with formulas such as 
(\ref{100}) one
would presumably
take then $T_1=\Lambda$.  It follows first in \cite{By} that $\beta_{ij}$
in (\ref{304}) plays the role of a metric $\eta_{ij}$ for WDVV equations
in the sense that (write ${\cal F}$ now for $F_{SW}$)
$$
{\cal F}_{ik\ell}\beta^{\ell m}{\cal F}_{mnj}=
{\cal F}_{jk\ell}\beta^{\ell m}{\cal F}_{mni};\,\,\beta^{\ell m}=
\eta^{\ell m};
$$
\be
c^i_{jk}=\eta^{ip}c_{jkp};\,\,c_{ijk}=c^p_{ij}\eta_{pk};\,\,
c_{ijk}=\partial_i\partial_j\partial_k{\cal F}
\label{109}
\ee
With the addition of an $a_0$ variable this can be extended to the more
general WDVV form of \cite{mr,Mw,Ms,Mz} as
\be
{\cal F}_i{\cal F}_k^{-1}{\cal F}_j={\cal F}_j{\cal F}_k^{-1}{\cal F}_i
\label{306}
\ee
where $({\cal F}_k)_{ij}={\cal F}_{ijk}$ and $i,j,k$ are now in the range
$0\to N-1$.
We see moreover that going back to (\ref{109}) one can take ${\bf (O)}\,\,
\eta_{ij}=c_{1ij}=\partial_1\partial_i\partial_jF$ as long 
as $\partial_i\sim
\partial/\partial a_i$ and one recalls that such a form is canonical in
standard WDVV theory for Whitham hierarchies (cf. \cite{cw,cb,dc,db,
ka,kc}).  Also the relations (\ref{109}) as developed in \cite{By} do not
use the residue formulation for $F_{ijk}$ of \cite{mr,Mw,Ms,Mz} 
so (\ref{109})
has a genuine Whitham flavor.  On the other hand the residue formulation
of $F_{ijk}$ via differentials $d\omega_i,\,\,d\omega_j,\,\,
d\omega_k$ in \cite{mr,Mw,Ms,Mz} yields (\ref{306}) and it would be
interesting to see how this is related to (\ref{306}) in terms of a
Whitham formulation (more on this below).

\section{RELATIONS TO \cite{ea,ia,ka,kc,ko,na}}
\renewcommand{\theequation}{3.\arabic{equation}}\setcounter{equation}{0}

The formulation in Section 2, based on \cite{gf}, differs from
\cite{ea,ia,ka,kc,ko,na} in certain respects and we want to clarify
the connections here.  Thus first we sketch very briefly some of the
development in \cite{na} (cf. also \cite{bb,cc}).  
Toda wave functions with a discrete parameter $n$ lead via $T_k=\epsilon
t_k,\,\,\bar{T}_k=\epsilon \bar{t}_k,\,\,T_0=-\epsilon n$, and $a_j
=i\epsilon\theta_j\,\,(=\oint_{A_j}dS)$ to a quasiclassical 
(or averaged) situation where (note $\bar{T}$ does not mean complex
conjugate)
\be
dS=\sum_1^ga_id\omega_i+\sum_{n\geq 0}T_nd\Omega_n+\sum_{n\geq 1}\bar{T}_n
d\bar{\Omega}_n
\label{1}
\ee
\be
F=\frac{1}{2}\left(\sum_1^ga_j\frac{\partial F}{\partial a_j}+
\sum_{n\geq 0}T_n\frac{\partial F}{\partial T_n}+\sum_{n\geq 1}
\bar{T}_n\frac{\partial F}{\partial \bar{T}_n}\right)
\label{2}
\ee
where $d\Omega_n\sim d\Omega_n^{+},\,\,d\bar{\Omega}_n\sim d\Omega_n^{-},
\,\,\bar{T}_n\sim T_{-n},$ and near $P_{+}$
\be
d\Omega_n^{+}=\left[-nz^{-n-1}-\sum_1^{\infty}q_{mn}z^{m-1}\right]dz\,\,
(n\geq 1);
\label{130}
\ee
$$d\Omega^{-}_n=\left[\delta_{n0}z^{-1}-\sum_1^{\infty}r_{mn}z^{m-1}\right]
dz\,\,(n\geq 0)$$
while near $P_{-}$
\be
d\Omega^{+}_n=\left[-\delta_{n0}z^{-1}-\sum_1^{\infty}\bar{r}_{mn}z^{m-1}
\right]dz\,\,(n\geq 0);
\label{131}
\ee
$$d\Omega^{-}_n=\left[-nz^{-n-1}-\sum_1^{\infty}\bar{q}_{mn}z^{m-1}\right]
dz\,\,(n\geq 1)$$
Here $d\Omega_0$ has simple poles at $P_{\pm}$ with residues $\pm 1$ and
is holomorphic elswhere; further $d\Omega_0^{+}=d\Omega_0^{-}=d\Omega_0$
is stipulated.  In addition the Abelian differentials 
$d\Omega_n^{\pm}$ for $n\geq 0$
are normalized to have zero $A_j$ periods and for the
holomorphic differentials $d\omega_j$ we write at $P_{\pm}$ respectively
\be
d\omega_j=-\sum_{m\geq 1}\sigma_{jm}z^{m-1}dz;\,\,d\omega_j=
-\sum_{m\geq 1}
\bar{\sigma}_{jm}z^{m-1}dz
\label{3}
\ee
where $z$ is a local coordinate at $P_{\pm}$.
Further for the SW
situation where ($g=N-1$)
\be
dS=\frac{\lambda dP}{y}=\frac{\lambda P'd\lambda}{y};\,\,
y^2=P^2-\Lambda^{2N};
\,\,P(\lambda)=\lambda^N+\sum_0^{N-2}u_{N-k}\lambda^k
\label{4}
\ee
(cf. (\ref{58}) where the notation is slightly different)
one can write near $P_{\pm}$ respectively
\be
dS=\left(-\sum_{n\geq 1}nT_nz^{-n-1}+T_0z^{-1}-\sum_{n\geq 1}
\frac{\partial F}
{\partial T_n}z^{n-1}\right)dz;
\label{5}
\ee
$$dS=\left(-\sum_{n\geq 1}n\bar{T}_nz^{-n-1}-T_0z^{-1}-\sum_{n\geq 1}
\frac{\partial F}{\partial\bar{T}_n}z^{n-1}\right)dz$$
leading to
\be
F=\frac{1}{2}\left(\sum_1^{N-1}\frac{a_j}{2\pi i}
\oint_{B_j}dS-\sum_{n\geq 1}
T_nRes_{+}z^{-n}dS-\right.
\label{6}
\ee
$$-\left.\sum_{n\geq 1}\bar{T}_nRes_{-}z^{-n}dS-T_0
[Res_{+}log(z)dS-Res_{-}log(z)dS]\right)
$$
(the $2\pi i$ is awkward but let's keep it - note one defines 
$a_j^D=\oint_{B_j}dS$).
In the notation of \cite{na} one can write now 
$(\bullet)\,\,h=y+P,\,\,\tilde{h}
=-y+P,$ and $h\tilde{h}=\Lambda^{2N}$ with $h^{-1}\sim z^N$ at $P_{+}$
and $z^N\sim \tilde{h}^{-1}$ at $P_{-}$  
(evidently $h\sim w$ of Section 2).  Note also 
$(\bullet\bullet)\,\,h+(\Lambda^{2N}/h)
=2P$ and $2y=h-(\Lambda^{2N}/h)$ yielding $y^2=P^2-\Lambda^{2N}$ and
calculations in \cite{na} give $(\bullet\bullet\bullet)\,\,dS=
\lambda dP/y=\lambda dy/P=\lambda dh/h$ (so $h\sim w$ in Section 2). 
Further the holomorphic
$d\omega_i$ can be written as linear combinations of holomorphic
differentials ($g=N-1$)
\be
dv_k=\frac{\lambda^{k-1}d\lambda}{y}\,\,(k=1,\cdots,g);\,\,
y^2=P^2-1=\prod_1^
{2g+2}(\lambda-\lambda_{\alpha})
\label{121}
\ee
(cf. (\ref{63}) where the notation
differs slightly).  Note also from (\ref{121}) that
$
2ydy=\sum_1^{2N}\prod_{\alpha\not=\beta}(\lambda-\lambda_{\alpha})d\lambda$
so $d\lambda=0$ corresponds to $y=0$.
From the theory of \cite{na} (cf. also \cite{cc}) one has then Whitham
equations
\be
\frac{\partial d\omega_j}{\partial a_i}=\frac{\partial 
d\omega_i}{\partial a_j};
\,\,\frac{\partial d\omega_i}{\partial T_A}=
\frac{\partial d\Omega_A}{\partial a_i};\,\,\frac{\partial d\Omega_B}
{\partial T_A}=\frac{\partial d\Omega_A}{\partial T_B}
\label{8}
\ee
along with structural equations
\be
\frac{\partial dS}{\partial a_i}=d\omega_i;\,\,\frac{\partial dS}{\partial
T_n}=d\Omega_n^{+};\,\,\frac{\partial dS}{\partial 
\bar{T}_n}=d\Omega_n^{-};
\,\,\frac{\partial dS}{\partial T_0}=d\Omega_0
\label{9}
\ee
Finally we note that in \cite{na} one presents a case (cf. also \cite{cb})
for identifying $N=2$ susy Yang-Mills (SYM) with a coupled system of
two topological string models based on the $A_{N-1}$ string.  This seems
to be related to the idea of $t\bar{t}$ fusion (cf. \cite{cd,db,ga}).
\\[3mm]\indent
Next we refer to the Krichever formulation $dS=QdE$ (cf. \cite{kc}).
Thus going to \cite{ko}
one takes a RS $\Sigma_g$ of genus $g$ with $M$ punctures $P_{\alpha}$.
Pick Abelian differentials $dE$ and $dQ$ such that $E$ and $Q$ have poles
of order $n_{\alpha}$ and $m_{\alpha}$ respectively at $P_{\alpha}$ and set
$dS=QdE$ with a pole of order $n_{\alpha}+m_{\alpha}+1$ at $P_{\alpha}$.
Pick local coordinates $z_{\alpha}$
near $P_{\alpha}$ so that $E\sim z_{\alpha}^{-n_{\alpha}}+R^E_{\alpha}log
(z_{\alpha})$, require $\oint_{A_j}dQ=0$, and fix the additive constant
in $S$ by requiring that its expansion near $P_1$ have no constant
term.  Define times
\be
T_{\alpha,i}=-\frac{1}{i}Res_{P_{\alpha}}(z_{\alpha}^idS)\,\,
(1\leq \alpha\leq M,\,\,1\leq i\leq n_{\alpha}+m_{\alpha});
\,\,R_{\alpha}^S=Res_{P_{\alpha}}(dS)
\label{WJ}
\ee
where $2\leq \alpha\leq M$ in the last set.
This gives $\sum_1^M(n_{\alpha}+m_{\alpha})+M-1$ parameters.  The remaining
parameters needed to parametrize 
the space ${\cal M}_g(n,m)$ of the creatures indicated consist of the
$2M-2$ residues of $dE$ and $dQ$, namely $R^E_{\alpha}=Res_{P_{\alpha}}dE$
and $R^Q_{\alpha}=Res_{P_{\alpha}}dQ\,\,(2\leq \alpha\leq M)$, plus
$5g$ parameters
\be
\tau_{A_i,E}=\oint_{A_i}dE;\,\,\tau_{B_i,E}=\oint_{B_i}dE;\,\,
\tau_{A_i,Q}=\oint_{A_i}dQ;\,\,\tau_{B_i,Q}=\oint_{B_i}
dQ;\,\,a_i=\oint_{A_i}QdE
\label{WK}
\ee
where $1\leq i\leq g$ in the last set.  Then it is proved in 
\cite{ka} that, if ${\cal D}$ is the open set in ${\cal M}_g(n,m)$ where
the zero divisors $\{z;\,dE(z)=0\}$ and $\{z;\,dQ(z)=0\}$ do not intersect,
then the joint level sets of the set of all parameters except the $a_i$
define a smooth $g$-dimensional foliation of ${\cal D}$.  Further near
each point in ${\cal D}$ the $5g-3+3M+\sum_1^M(n_{\alpha}+m_{\alpha})$
parameters $R^E_{\alpha},\,R^Q_{\alpha},\,R^S_{\alpha},\,T_{\alpha,k},\,
\tau_{A_i,E},\,\tau_{B_i,E},\,\tau_{A_i,Q},\,\tau_{B_i,Q},$ and $a_i$
have linearly independent differentials and thus define a local holomorphic
coordinate system.
Assume now that $dE$ has simple zeros $q_s\,\,(s=1,\cdots,2g+n-1$ in
the case of ${\cal M}_g(n,1)$ below)
and we come to the Whitham
times.
The idea here is that suitable submanifolds of ${\cal M}_g(n,m)$ are
parametrized by $2g+M-1+\sum_1^M(n_{\alpha}+m_{\alpha})$ Whitham
times $T_A$ to each of which is associated a dual time $T^D_A$ and
an Abelian differential $d\Omega_A$.  For $M=1$ (one puncture)
one has
\be
T_j=-\frac{1}{j}Res(z^jdS);\,\,T^D_j=Res(z^{-j}dS);\,\,d\Omega_j
\,\,(1\leq j\leq n+m)
\label{WS}
\ee
($d\Omega_j\sim d\Omega_j^{+}
=d(z^{-j}+O(z)$ near $P_1$) with $\oint_{A_j}d\Omega_i=0$).
For $g>0$ there are $5g$ more parameters and we consider only foliations
for which $\oint_{A_k}dE,\,\,\oint_{B_k}dE$, and $\oint_{A_k}dQ$ are fixed.
This leads to 
\be
a_k=\oint_{A_k}dS;\,\,T^E_k=\oint_{B_k}dQ;\,\,a^D_k=
-\frac{1}{2\pi i}\oint_{B_k}dS;\,\,{}^DT^E_k=\frac{1}{2\pi i}
\oint_{A_k^{-}}EdS
\label{WT}
\ee
The corresponding differentials are $d\omega_k$ and $d\Omega^E_k$ where 
the $d\Omega^E_k$ are holomorphic on $\Sigma$ except along $A_j$ cycles
where $(\clubsuit)\,\,d\Omega^{E_{+}}_k-d\Omega^{E_{-}}_k=
\delta_{jk}dE$.
We will assume that $Q$ is holomorphic away from the $P_{\alpha}$ and
thus jumps on the $A_j$ cycles are not entertained; hence
the variables $T_{Q,k}\sim T^Q_k$ of \cite{kc} will not arise.  
Thus one has $2g+n+m$ times $T_A=(T_j,a_k,T^E_k)$. 
For $M>1$ punctures there are $2g+\sum (n_{\alpha}+m_{\alpha})$ times
$(T_{\alpha,j},a_k,T^E_k)$ plus $3M-3$ additional parameters for the residues
of $dQ,\,dE,$ and $dS$ at the $P_{\alpha}\,\,(2\leq\alpha\leq M)$.
For convenience one considers only leaves where $(\spadesuit)
\,\,Res_{P_{\alpha}}dQ=0;\,\,Res_{P_{\alpha}}dE=\,\,fixed\,\,(2\leq\alpha
\leq M)$ and incorporates among the $T_A$ the residues $R_{\alpha}^S=
Res_{P_{\alpha}}dS\,\,(2\leq\alpha\leq M)$ with $M-1$ dual times
$(\clubsuit\clubsuit)\,\,{}^DR^S_{\alpha}=-\int_{P_1}^{P_{\alpha}}
dS$ where $2\leq\alpha\leq M$, corresponding to differentials
$d\Omega^3_{\alpha}$ which are Abelian differentials of third kind with 
simple poles at $P_1$ and $P_{\alpha}$ and residue $1$ at $P_{\alpha}$.  The
Whitham tau function is $\tau=exp({\cal F}(T))$ where
\be
{\cal F}(T)=\frac{1}{2}\sum_AT_AT^D_A+\frac{1}{4\pi i}\sum_1^ga_k
T^E_kE(A_k\cap B_k)
\label{WU}
\ee
Here $A_k\cap B_k$ is the point of intersection of these cycles.  When
$Res_{P_{\alpha}}dE=0$ one obtains the derivatives of ${\cal F}$ with
respect to the $2g+\sum(n_{\alpha}+m_{\alpha})+M-1$ Whitham times as
\be
\partial_{T_A}{\cal F}=T^D_A+\frac{1}{2\pi i}\sum_1^g\delta_{a_k,A}
T^E_kE(A_k\cap B_k);
\label{WV}
\ee
$$\partial^2_{T_{\alpha,i},T_{\beta,j}}{\cal F}
=Res_{P_{\alpha}}(z_{\alpha}^id\Omega_
{\beta,j});\,\,\partial^2_{a_j,A}{\cal F}=\frac{1}{2\pi i}
\left(E(A_k\cap B_k)
\delta_{(E,k),A}-\oint_{B_k}d\Omega_A\right);$$
$$\partial^2_{(E,k),A}{\cal F}=\frac{1}{2\pi i}\oint_{A_k}Ed\Omega_A;\,\,
\partial^3_{ABC}{\cal F}=\sum_{q_s}Res_{q_s}\left(\frac{d\Omega_A
d\Omega_Bd\Omega_C}{dEdQ}\right)$$
When $Res_{P_{\alpha}}dE\not= 0$ one has $(\spadesuit\spadesuit)\,\,
\partial{\cal F}/\partial R^S_{\alpha}={}^DT^S_{\alpha}+(1/2)\pi i
\sum c_{\alpha\beta}R^S_{\beta}$ with antisymmetric integer $c_{\alpha
\beta}$.
\\[3mm]\indent
For one puncture the case of interest would be $Q_{+}=z^{-1}$ and there
are two Whitham times $T_n=0$ and $T_{n+1}=n/(n+1)$ fixed so we will
have $2g+n-1$ Whitham times for ${\cal M}_g(n,1)$.
Next one shows that each $2g+n-1$ dimensional leaf $\hat{{\cal M}}$ of the 
foliation of ${\cal M}_g(n)$ parametrizes the marginal deformation of a
TFT on $\Sigma$.  The free energy of such theories is the restriction
of ${\cal F}$
to the appropriate leaf.  Thus we consider the leaf within ${\cal M}_g(n,1)$
of dimension $2g+n-1$ which is defined by the constraints
\be
T_n=0;\,\,T_{n+1}=\frac{n}{n+1};\,\,\oint_{A_k}dE=0;\,\,\oint_{A_k}dQ=0;\,\,
\oint_{B_k}dE= fixed
\label{111}
\ee
Thus the leaf is parametrized by the $n-1$ Whitham times $T_A\,\,(A=1,
\cdots,n-1)$ and by the periods $a_k=\oint_{A_k}dS$ and 
$T^E_k=\oint_{B_k}dQ$.  There will be primary fields $\phi_i\sim
d\Omega_i/dQ\,\,(i=1,\cdots,n-1)$ plus $2g$ additional fields
$d\omega_i/dQ$ and $d\Omega^E_j/dQ$.  We will use the symbol $\sim$
to mean either ``is asymptotic to", or ``corresponds to", or ``is
associated with"; the meaning should be clear from context. 
Then one can define
\be
\eta_{A,B}=\sum_{q_s}Res_{q_s}\frac{d\Omega_Ad\Omega_B}{dE};\,\,
c_{ABC}=\sum_{q_s}Res_{q_s}\frac{d\Omega_Ad\Omega_Bd\Omega_C}{dEdQ}
=F_{ABC}
\label{112}
\ee
where $T_A\sim (T_i,a_j,T^E_k)$.  
The formulas (\ref{WV}) hold as before and the Whitham equations 
are generically 
$\partial_Ad\Omega_B=\partial_Bd\Omega_A$ which can in
fact be deduced from $\partial_AE=\{\Omega_A,E\}$ where $\{f,g\}=f_pg_X
-g_pf_X$ with $dp\sim d\Omega_1$ (cf. \cite{cc,ka,kc,ko}).
We see that for $A=1,\,\,d\Omega_A=dQ$ implies formally $c_{1BC}=\eta_{BC}$
so $T_1$ plays a special role in the general theory with one puncture.
\\[3mm]\indent
For two punctures we follow now \cite{cb,dz,dg,ka,kc,ko} with $N_c=N,\,\,
N_f=0,$ and no masses $m_i$.  Thus $\lambda_{SW}\sim dS=QdE$ and for 
$SU(N)$ theories 
${\bf (P)}\,\,
dE$ has simple poles at points $P_{\pm}$ with residues $-N$ and
$N$.  Its periods around homology
cycles are multiples of $2\pi i.\,\,
{\bf (R)}\,\,Q$ is a meromorphic function with simple poles 
only at $P_{\pm}$.
The other parameters of the leaf are determined by the
following normalizations of $dS=QdE$
\be
Res_{P_{+}}(zdS)=-N2^{-1/N};\,\,
Res_{P_{-}}(zdS)=N\Lambda^22^{-1/N};\,\,
Res_{P_{+}}(dS)=0
\label{ff}
\ee
These conditions imply that $\Sigma$ is hyperelliptic
and has an equation of the form
\be
y^2=\prod_1^N(Q-\bar{a}_k)^2-\Lambda^{2N}
\equiv A(Q)^2-\Lambda^{2N};\,\,dS\sim\frac{Q}{y}dA
\label{g}
\ee
(cf. here (\ref{4})). There are
corrections $a_k=\bar{a}_k+O(\Lambda^N)$
which can be absorbed in a reparametrization
leaving ${\cal F}$ invariant so that one can identify the $\bar{a}_k$ of
(\ref{g}) with diagonal elements in a Cartan subalgebra decomposition
with $\sum_1^N\bar{a}_k=0$.  
If one represents the RS (\ref{g}) by a two
sheeted covering of the complex plane then $Q$ is just the coordinate
in each sheet while $(\clubsuit\clubsuit\clubsuit)\,\,
E=log(y+A(Q))$.  
The points $P_{\pm}$ are points at
infinity with the two possible sign choices $\pm$ for $y=\pm\sqrt
{A^2-B}$.  The
prepotential then satisfies ($z$ is a local coordinate)
\be
\sum_1^{N_c}a_j\frac{\partial{\cal F}}{\partial a_j}
-2{\cal F}= {\cal D}{\cal F}=
\label{h}
\ee
$$=-\frac{1}{2\pi i}
\left[Res_{P_{+}}(zdS)Res_{P_{+}}(z^{-1}dS)
+Res_{P_{-}}(zdS)Res_{P_{-}}(z^{-1}dS)\right]$$
It is known that the right side of
(\ref{h}) is a modular form (cf. \cite{bc,ba,be}) and one arrives at
$(\spadesuit\spadesuit\spadesuit)\,\,
{\cal D}{\cal F}=-(N/2\pi i)\sum_1^N\bar{a}_k^2$.
Thus near $P_{+}$ 
one writes $y^2=\prod_1^N(Q-\bar{a}_k)^2-\Lambda^{2N}=P^2
-\Lambda^{2N}$ (so $Q\sim \lambda$ in a sense indicated also in
(\ref{g})) and $E=
log(y+P)$ which means $E\sim log(h)$ with $dE=dh/h$ as
above.  Note also e.g. $z^N\sim h^{-1}$ at $P_{+}$ corresponds to $Nlog(z)
=-log(h)\sim -E$ or $E=log(h)$.  Further near $P_{+}$
\be
E\sim -Nlog(z);\,\,Q\sim 2^{-1/N}z^{-1}+O(1)
\label{133}
\ee
so $E=Nlog(Q)+log(2)+O(Q^{-1})$,
while near $P_{-},\,\,Q\sim (\Lambda/2)^{1/N}z^{-1} +O(1)$ with
\be
E=-Nlog(Q)+log(2)+log\left(\frac{\Lambda}{4}\right)+O(Q^{-1})
\label{134}
\ee
Thus $dE\sim -Nd\Omega_0$.
\\[3mm]\indent {\bf REMARK 3.1.}$\,\,$
Note that the
formats of \cite{dc,kc} do not include $log(Q)$ terms in $E$ but we
record here a few facts from \cite{dc,kc} for background
and to illustrate some examples.  Thus first, Hurwitz spaces
are moduli spaces of RS of a given genus $g$ with $n+1$ sheets, i.e.
of pairs $(\Sigma_g,E)$ where $E$ is a meromorphic function on $\Sigma_g$
of degree $n+1$.  The ramification is to be entirely determined by $E$.  One
assumes there are $m+1$ punctures $\infty_j$ and that $E$ has degree
$n_j+1$ at $\infty_j\,\,(E:\,\Sigma_g\to {\bf CP^1}$ and $E^{-1}(\infty)=
\infty_0\cup\cdots\cup \infty_m$) with $n=2g+n_0+\cdots+n_m+2m$.  Thus
for one puncture $n=2g+n_0$.  The ramification points of $\Sigma_g$ are
$u^j=E(P_j)$ where $dE(P_j)=0\,\,(j=1,\cdots,n)$ and the $P_j$ are the
branch points of $\Sigma_g$.  To see how this works we consider
${\bf (S)}\,\,g=0,\,m=0,\,n_0=n$ with $E=k^{n+1}+a_nk^{n-1}+\cdots+a_1$
where $k$ represents a local coordinate at $\infty_0=\infty$.  It is 
important to note here that if $\Sigma_g$ is the RS of an irreducible
algebraic equation $P(z,k)=0$ of degree $N$ in $k$ and if the $r$ branch 
points have orders $N_i\,\,(N_i+1\sim$ degree) then the Riemann-Hurwitz
(RH) formula says $g=1-N+(1/2)\sum_1^rN_i$ (cf. \cite{ja}).  Thus
in ${\bf (S)}$ if $E'(k)=\prod_1^n(k-k_j)$ with distinct $k_j$ then $N_j=1$
and there is a branch point at $\infty$ of order $n$.  Hence for
$P(E,k)=E-k^{n+1}-a_nk^{n-1}-\cdots-a_1=0$ one has
$g=1-(n+1)+(1/2)\sum_1^n1+(1/2)n=0$ and we see how an $N$ sheeted surface
can have genus zero.  Another example is ${\bf (T)}\,\,
g>0,\,m=0,\,n_0=1$ with
hyperelliptic curves $k^2=\prod_1^{2g+1}(E-E_j)$ with 
branch points $E_j$ and
$\infty$.  Note $2kdk=[\sum\prod'(E-E_j)]dE$ so $dE=0$ where $k=0$, i.e.
at $E_j$.  The RH formula gives correctly $g=1-2+(1/2)
\sum_1^{2g+1}1+(1/2)$
while $n=2g+1$.
Another special example is ${\bf (U)}\,\,$ namely elliptic curves
$\mu^2=4(E-c)^3-g_2(E-c)-g_3=4(E-c-e_1)(E-c-e_2)
(E-c-e_3)$.  Using Weierstrass uniformization one writes 
$E={\cal P}(z)+c$ and
$k={\cal P}'(z)$ where ${\cal P}={\cal P}(z,g_2,g_3)$ is the Weierstrass
function.  The infinity point is $z=0$ and one can take ${\cal P}(\omega)=
e_1$ with ${\cal P}(\omega')=e_3$.  Setting $dk=dz/2\omega$ one obtains
$E(k)={\cal P}(2\omega k,\omega,\omega')+c$ with $k\simeq k+m+n\tau\,\,
(\tau=\omega'/\omega)$.
\\[3mm]\indent
Now for two punctures we look first at the moduli space $M_{g,n_{\pm}}$ where
$n=2g+n_{+}+n_{-}$ and $E^{-1}(\infty)=\infty_{+}\cup\infty_{-}$ (here
$n_{+}\sim n_0$ and $n_{-}\sim n_1$).  We are going to want $E$ as in
(\ref{133}) - (\ref{134}) corresponding to $dP/y$ with $P$ as in 
(\ref{4}) and $dS\sim QdE$ will be used with $Q\sim \lambda$ as indicated
in (\ref{4}) and (\ref{g}).  Thus $dE(q_s)=0$ for distinct simple zeros
$q_s$ means $\#(q_s)-2=2g-2$ by Riemann-Roch (cf. 
\cite{sc}) so $s=1,\cdots,2g$
(recall this is $2g-1$ for one puncture in ${\cal M}_g(n,1)$ with $n=0$).
Now looking at the differentials and times in \cite{dc} we see that the
situation (\ref{133}) - (\ref{134}) is not covered ($log(p)$ terms are
not included).  The development of Hurwitz spaces could clearly be expanded
to cover this situation but we will not dwell on this here.
The formulation of \cite{kc} is similar in that no logarithmic terms are
involved in $E$.
Thus go to \cite{dz,ka,ko} and
look at the two puncture situation $P_{\pm}$ with special attention
to $N=2$ susy YM (SW theory).  Thus take $P_1\sim P_{+}$ and 
$P_2\sim P_{-}$ so by (\ref{133}) - (\ref{134}) at $P_{+},\,\,
E\sim -Nlog(z)$ 
and $Q\sim 2^{-1/N}z^{-1}$ while at $P_{-},\,\,E\sim Nlog(z)$ and 
$Q\sim 2^{-1/N}\Lambda^{1/N}z^{-1}$.  This means $R^E_{+}=-N$ and $R^E_{-}
=N$ while $R^Q_{+}=R^Q_{-}=0$ since $Res_{\pm}dQ=0$.  The $T_{\alpha,i}$ 
have the form
\be
T^{+}_1=-Res_{+}(zdS)=N2^{-1/N};\,\,T^{-}_1=-Res_{-}(zdS)=-N2^{-1/N}
\Lambda^{1/N}
\label{10}
\ee
while $R^S_{-}=Res_{-}dS=0$
($n_{\pm}=0$ and $M_{\pm}=1$ implies $1\leq i\leq 1$ in (\ref{WJ})).  To
see this note that $dS=QdE\sim -N2^{-1/N}z^{-2}dz$ near $P_{+}$ and $dS\sim
N2^{-1/N}\Lambda^{1/N}z^{-2}dz$ near $P_{-}$.  Note also $M=2$ so
$(n_{\pm}+m_{\pm}+2-1=3$ for the parameters $T^{\pm}_1$ and $R^S_{-}$.
Further $R^E_{-}=Res_{-}dE=N$ and $R^Q_{-}=0$ gives $2=2M-2$ parameters.
The $5g$ parameters in (\ref{WK}) are all present a priori.  Recall that
the idea is to parametrize suitable submanifolds of ${\cal M}_g(n,m)$ by
$2g+2-1+n_{\pm}+m_{\pm}=2g+1+2=2g+3$ Whitham times $T_A$.
One will have $2g+2$ times $(T^{\pm}_1,a_k,T^E_k)$ (cf. (\ref{WT} for
$T^E_k=\oint_{B_k}dQ\sim \tau_{B_k,Q}$ in (\ref{WK})) and 3 additional
parameters $R^Q_{-}=0,\,\,R^E_{-}=N,$ and $R^S_{-}=0$ (note $-d\Omega_0$ is
associated with $R^S_{-}$ here so in some sense $R^S_{-}\sim T_0$
and this is indeed $0$ d'apr\`es \cite{ea}).
In \cite{ko} this set of times is referred to as Whitham times on leaves
of ${\cal M}_g(n,m)$ where $\tau_{A_kE}=\oint_{A_k}dE,\,\,\tau_{B_k,E}=
\oint_{B_k}dE,$ and $\tau_{A_k,Q}=\oint_{A_k}dQ$ are fixed.  Thus in
particular one is not using times $T^{\pm}_j$ associated to differentials
$d\Omega^{\pm}_j$ for $j>1$ which seems to be saying that in the two
puncture situation Whitham theory on certain prescribed leaves of 
${\cal M}_g(n,m)$ involves only $2g+3$ times as indicated (since such leaves
are parametrized by such a set of times).  Evidently other 
times $T^{\pm}_j$
will not occur under the definitions since if we write in (\ref{WJ})
$dS=QdE\sim [(c^{\pm}/z)+\sum_0^{\infty}q^{\pm}_nz^n]
[\mp (N/z)+\sum_0^{\infty}
e^{\pm}_nz^n]dz$ then $Res_{\pm}z^jdS=0$ for $j>1$.  
This is consistent with the development above
and quite different
from LG situations coupled to gravity
or generic Whitham situations as in \cite{Aa,cc,ch,
ci,cj,dc,kc,lb,ta,tc} where an infinity of times $T_A$ arise naturally
(also in \cite{gf,na} for example before specialization to SW theory).
Note also in (\ref{h}) that the right side will involve $Res_{+}z^{-1}dS
\sim {}^DT^{+}_1=c^{+}e_1^{+}-Nq^{+}_1+q_0^{+}e_0^{+}$ and $Res_{-}z^{-1}
dS\sim {}^DT_1^{-}=c^{-}e_1^{-}+Nq_1^{-}+q_0^{-}e_0^{-}$ which shows
that the variables $T_1^{\pm}$ (and their duals) can be used to restore
homogeneity in ${\cal F}$ (this differs slightly from \cite{ea} where the
$T^{-}_j$ are not used).
Note that in \cite{ea} (and also in \cite{gc,ia}) this point of view
is expressed by saying that when the higher $T_n$ are zero in a full
Whitham theory (as in \cite{cc,na}) and $T_0,\,T^{\pm}_1$ are given
special values, one recovers the form of $dS$ and ${\cal F}$ appropriate
to SW theory.  This is accomplished directly in \cite{ka,ko} by the
definitions of $T^{\pm}_j$ in (\ref{WJ}) and the formulation on leaves 
of ${\cal M}_g(n,m)$.  One notes here that in \cite{na} (cf. also \cite{ea})
\be
\frac{1}{2\pi i}\frac{\partial{\cal F}}{\partial T_n}=-Res_{+}z^{-n}dS
=-{}^DT^{+}_n;\,\,\frac{1}{2\pi i}\frac{\partial{\cal F}}{\partial\bar{T}_n}
=-Res_{-}z^{-n}dS=-{}^DT^{-}_n
\label{11}
\ee
(we have inserted the factor $1/2\pi i$) so the right side
of (\ref{h}) is $-T^{+}_1(\partial{\cal F}/\partial T^{+}_1)-
T_1^{-}(\partial{\cal F}/\partial T_1^{-})$.  
\\[3mm]\indent {\bf SUMMARY 3.2.}$\,\,$  Thus (\ref{h}) says
(upon introducing the renormalization term 
$\Lambda\partial_{\Lambda}{\cal F}$ - cf. \cite{bc,By,ba,be,cb,ea})
\be
\sum a_j\frac{\partial{\cal F}}{\partial a_j}-2{\cal F}=-T_1^{+}\frac
{\partial{\cal F}}{\partial T_1^{+}}-T_1^{-}\frac{\partial{\cal F}}
{\partial T^{-}_1}=-\Lambda\frac{\partial{\cal F}}{\partial \Lambda}
\label{12}
\ee
This gives then the two puncture version of a result in \cite{ea} (cf.
\cite{cb}), namely, writing $T_1^{+}=X$ and $T_1^{-}=\bar{X}$
\be
\Lambda\partial_{\Lambda}{\cal F}=X\partial_X{\cal F}+\bar{X}\partial_
{\bar{X}}{\cal F}
\label{13}
\ee
One sees incidently that the definition of $T_j^{\pm}$ as in (\ref{WJ})
can be explained via formulas like (\ref{h}) which involves calculation
with Riemann bilinear relations.  Note that in making the explicit choices
of $T_1^{\pm}$ which identify $dS$ and ${\cal F}$ with a SW theory one
introduces a relation between $T_1^{-}$ and $\Lambda$ in (\ref{10}) of
the form $(\bullet\clubsuit\bullet)\,\,(T_1^{-})^N=c\Lambda$ (in \cite{ia}
with one puncture, using the formulation of \cite{gc} one obtains
$T_1=(\sqrt{2}/\pi)\Lambda$ - it is curious that this does not appear
in the more canonical formulas of \cite{ea} but we note that $\Lambda$ is not
involved in $T_1^{+}$ in (\ref{10})).  Note also for $\bar{X}^N=c\Lambda$
one has $N(\partial\bar{X}/\partial\Lambda)\bar{X}^{N-1}=c$ 
and $\partial{\cal F}/
\partial\Lambda=(\partial{\cal F}/\partial\bar{X})(c/N)(1/\bar{X}^{N-1})$
which implies that $\Lambda(\partial{\cal F}/\partial\Lambda)=(1/N)\bar{X}
(\partial{\cal F}/\partial\bar{X})$.  This introduces the possible 
identification $\bar{X}^N=c\Lambda$ in the theory (instead of $T_1=
(\sqrt{2}/\pi)\Lambda$).  
The two puncture situation as described suffers however from
a lack of symmetry in (\ref{10}) with respect to $\Lambda$ and
this could surely be symmetrized to make a nice combination of 
$X$ and $\bar{X}$ correspond to $\Lambda$.  Indeed this is precisely
what has been attained in \cite{gf} by dealing with $d\Omega_j$
and $d\hat{\Omega}_j$ in a symmetric manner,
as indicated in Conclusion 2.1 (cf. also Proposition 3.3).
\\[3mm]\indent {\bf REMARK 3.3.}$\,\,$
With this background let us think of a general Whitham theory in the spirit
of \cite{ka,kc,ko} for the RS (\ref{58}) - (\ref{60}) or the equivalent
forms of \cite{na} or \cite{ko} with basic differentials 
$d\Omega_n,\,\,d\hat{\Omega}_n\sim
T_n$ as in Section 2 for $n\geq 1$, plus $d\Omega_0\sim T_0\sim R^S_{-}$,
plus the holomorphic differentials $d\omega_j$. 
Set further, instead of $dE$ as before, $d{\cal E}=\alpha(dw/w)$
(which corresponds to $-\alpha dE\sim\alpha N\,d\Omega_0)$) and
consider $\hat{\Omega}_1\sim \beta w^{\alpha}$ with $d\hat{\Omega}_1
=\hat{\Omega}_1d{\cal E}$ a tautology.  Recall $dS_{SW}=d\hat{\Omega}_1
=P_{+}^{1/N}(dw/w)=\lambda(dw/w)$ and $\lambda\sim w^{\pm 1/N}$ near
$P_{\pm}$ from Section 2.  Now near $P_{+}$ for example we consider
$\beta dw^{\alpha}=\beta\alpha w^{\alpha-1}dw=\lambda (dw/w)=
w^{(1/N)-1}dw$ to conclude that $\alpha=1/N$ and $\beta= 1/\alpha = N$.
Near $P_{-}$ one has $\lambda\sim w^{-1/N}$ with $\beta\alpha 
w^{\alpha -1}=w^{-(1/N)-1}dw$ or $\alpha=-1/N$ and $\beta=-N$, which
simply means that $\hat{\Omega}_1\sim\pm Nw^{\pm 1/N}\sim \pm N\lambda$
near $P_{\pm}$.  Now recall $d\Omega_1$ has the same singularity
behavior as $d\hat{\Omega}_1$
near $\infty_{\pm}\sim 0=
\xi= w^{\mp 1/N}$,
plus a normalization $\oint_{A_i}d\Omega_1=0$.
Then,
setting $\hat{\Omega}_1=\int d\hat{\Omega}_1\sim
\pm Nw^{\pm 1/N}$ near $P_{\pm}$ with $d{\cal E}=(1/N)(dw/w)\sim d\Omega_0$
one has $\hat{\Omega}_1d{\cal E}=\lambda(dw/w)=dS_{SW}$ and the formalism
suggests that we should have formulas (cf. (\ref{112}))
\be
c_{ABC}=\sum_{q_s}Res\frac{d\Omega_Ad\Omega_Bd
\Omega_C}{d\hat{\Omega}_1d{\cal E}}=\frac{\partial^3F}
{\partial T_A\partial T_B\partial T_C};
\label{308}
\ee
As for $\eta$ one would then examine heuristically
\be
\eta_{AB}=c_{1AB}=\sum_{q_s}Res\frac{d\Omega_A
d\Omega_Bd\Omega_1}{d\hat{\Omega}_1d{\cal E}}
\label{500}
\ee
where $d{\cal E}(q_s)=0$
and these formulas should apply to the basic $d\Omega_n\,\,(n\geq 1),
\,\,d\Omega_0$, and the $d\omega_j$. 
The formulas in (\ref{308}) are heuristic but should simply correspond
to a reshuffling of terms
in (\ref{112}).  
\\[3mm]\indent {\bf REMARK 3.4.}$\,\,$
Given (\ref{308}) and (\ref{500})
one would have for $A,B\sim a_i,a_j$ the formula $\eta_{ij}=
\partial^3F_W/\partial\alpha_i\partial\alpha_j\partial T_1=c_{1ij}
=\sum_{q_s}Res[d\omega_id\Omega_1d\omega_j/{d\hat{\Omega}_1d\cal E}]$
and this could be nonzero since $d\hat{\Omega}_1$ should have
some zeros different from those of $d\Omega_1$ and the residue
sum would not revert to the zero residue at $\infty_{\pm}$; this is in
contrast to \cite{ko} where $\eta_{ij}=0$ in this situation for
one puncture but gives us a chance to compare with (\ref{304}).  Indeed 
one would like now to show that $\beta_{ij}=\eta_{ij}
=\Lambda\partial_{\Lambda}\tau_{ij}$ as in (\ref{304}) if we think
of $\Lambda\sim T_1$.  This seems like a resolvable problem but
we only examine a few details below (it is also connected to a WDVV
problem as indicated after (\ref{126})).  
Recall also that (\ref{500}) is still
only conjectural and furthermore
there is a certain
subtlety involved relative to $\Lambda$ and $T_1$ derivatives
(cf. (\ref{203}) and \cite{gf}).  Thus $F_{SW}$ contains no $T_n$
but $\Lambda\sim T_1$ is a permitted agreement.  On the other hand
$F_W=F_{Whit}$ as in (\ref{308}) contains both $\Lambda$ and $T_1$
with $\alpha_i=T_1a_i+O(T_2,T_3,\cdots)$ so $\Lambda$ and $T_1$ derivatives
will both occur and 
will be related in some manner similar to (\ref{203}) (recall $F_W$ is 
homogeneous of degree two in the $T_n$ and $a_i$). 
Consider now residues at infinity in $\eta_{AB}$ for example
(which will not be enough to calculate $\eta_{AB}$ since the sum
of residues over $q_s$ and $\pm\infty$ do not exhaust all residues
of the differential in (\ref{500})).  From
(\ref{133}) - (\ref{134})
$d{\cal E}$ contributes a multiplier $\pm z$ via ${\cal E}\sim\pm
log(z)$.  Since the coefficients of the singular terms in $d\Omega_n$
are the same at $P_{\pm}$ with $w^{\pm n/N}\sim\xi^{-n}\,\,(z\sim\xi)$
in order to balance in (\ref{73}) 
we can write at $P_{\pm}$ for $n\geq 1$
\be
d\Omega_n=\left(\xi^{-n-1}+\sum_1^{\infty}s^{\pm}_{pn}\xi^{p-1}
\right)d\xi;\,\,\frac{d\Omega_1}{d\hat{\Omega}_1}=\sum_1^{\infty}
v^{\pm}_p\xi^{p-1}
\label{309}
\ee
(note this is consistent with (\ref{130}) - (\ref{131})).  Then
\be
\eta_{AB}^{\pm}=-Res_{\pm}\left(\pm\xi\frac{d\Omega_n
d\Omega_md\Omega_1}{d\hat{\Omega}_1d\xi}\right)=
\label{310}
\ee
$$=\mp Res_{\pm}\left[
\xi\left(\xi^{-n-1}+\sum_1^{\infty}s^{\pm}_{pn}\xi^{p-1}
\right)\left(
\xi^{-m-1}+\sum_1^{\infty}s^{\pm}_{\ell m}\xi^{\ell -1}\right)
\sum_1^{\infty}v^{\pm}_s\xi^{s-1}\right]d\xi$$
and only $T_1$ would be involved in a putative WDVV metric.
We recall also that in standard LG models where
$dE\sim \xi^{-n-1}$ one has $\eta_{AB}=\delta_{i+j,n}$ corresponding to
$A\sim T_i$ and $B\sim T_j$ with $i,j\leq n$ (cf. \cite{dc,ka,kc,ko,lb}).
We note also
a peculiarity here relative to \cite{ka,kc,ko}.  Thus 
recall $d\Omega_k^{{\cal E}}$ is holomorphic
except on $A_j$ cycles where $d\Omega_k^{{\cal E}_{+}}-
d\Omega_k^{{\cal E}_{-}}=-\delta_{jk}d{\cal E}$ (cf. $(\clubsuit)$).
However $T_k^{{\cal E}}\sim\oint_{B_k}dQ$ in 
(\ref{WT}) and this seems to be a problem here if $dQ\sim d\hat{\Omega}_1
\sim dS_{SW}$ since $(-1/2\pi i)\oint_{B_k}dS_{SW}=a_k^D$ in (\ref{WT}).
The results of \cite{kc} (where no logarithmic terms are involved)
suggest that for $A\sim a_j$ and $B\sim T_k^{{\cal E}}$ there results
$\eta_{AB}=\delta_{jk}$.  If that holds here and if $T_k^{{\cal E}}\sim 
ca_k^D$
it may indicate some kind of interplay between geometry and ``duality"
and/or the role of $T^{{\cal E}}_k$ as a ``basic" variable may
change in the present format.
In terms of possible gravitational couplings
for additional $T_j$, or simply deformations of a basic SW theory 
therewith, one does not expect the WDVV metric to extend but the 
$a_j$ and $T_1\sim\Lambda\sim a_0$ would become functions of the $T_j$
for $j>1$.  One could then imagine hydrodynamic type equations for 
$a_j(T_m),
\,\,0\leq j\leq g,\,\,m>1$ as in \cite{dc}.
\\[3mm]\indent {\bf REMARK 3.5.}$\,\,$
In an earlier version of this paper we wrote (\ref{308}) with
differentials $d\hat{\Omega}_A\sim \hat{T}_A$ for some undefined $\hat{T}$
(presumably meant to be $T_A$) in order to produce a ``clean"
formula for $\eta_{AB}=c_{1AB}$.  Since that resulted in $\eta_{ij}
=\eta_{AB}=0$ for $A,B\sim a_i,\,a_j$, which 
seems incompatible with (\ref{304})
if $\partial_{\Lambda}\sim\partial/\partial T_1$, we prefer the present
formulation which is also closer in spirit to \cite{ka,kc,ko}.
\\[3mm]\indent {\bf REMARK 3.6.}$\,\,$
This
should all be examined further in connection with special geometry
(cf. \cite{bd,cd,cn,fg,Ga,ld,wa}). In this spirit turning on
the Whitham dynamics could perhaps correspond to 
allowing deformations via $T_j$ for $j>1$.
Such terms $T_j$ are of course intrinsic to the RS via the BA function,
as indicated in Section 2.  The dependence of moduli (such as branch
points) on higher $T_j$ seems to indicate change of complex structure and
this corresponds to changing the Casimir moduli $h_k$ of the associated
Toda theory.  Given the independence of $a_j$ and $T_j$ as provided
for in \cite{ia} for example one can perhaps think of the $a_j$ variables
as determining the K\"ahler structure and thus variations in K\"ahler
structure and complex structure can be separated.  In that spirit the
renormalization parameter $\Lambda\sim a_0$ is naturally associated with
the K\"ahler structure in keeping with comments in \cite{fg}.

\section{WDVV}
\renewcommand{\theequation}{4.\arabic{equation}}\setcounter{equation}{0}

Now one goal of this paper is to understand the relations between WDVV
theory as in \cite{ka,ko} (modified via Section 2 to suitable
formulas such as (\ref{308}))
and the WDVV theory for the
$a_j$ alone as defined in \cite{By,ba} or in \cite{mr,Mw,Ms,Mz}  
(cf. also \cite{Kz} for still another approach).  We will
have to deal here with the variables $T^E_k=\oint_{B_k}dQ$ and 
$d\Omega^E_k$
(described in $(\clubsuit)$ via $d\Omega_k^{E_{+}}-d\Omega_k^{E_{-}}=
\delta_{jk}dE$ where $d\Omega^E_k$ is holomorphic except on $A_j$ cycles
as indicated).  Formulas of the form (\ref{WU}), (\ref{WV}), and (\ref{112})
are for one puncture however as are conclusions such as Theorem 18 in
\cite{ko} so we will have to find the two puncture version (cf. \cite{kc}
for a multi-puncture version for LG type theories without $log(Q)$ terms).
First we remark that
for TFT with a LG potential $W(\lambda)$ one writes
$\phi_i\phi_j=c^k_{ij}\phi_k\,\,mod\,W'$ with $F_{ijk}=Res[\phi_i\phi_j
\phi_k/W']=\sum[(\phi_i\phi_j\phi_k)(\lambda_{\alpha})/W''
(\lambda_{\alpha})]$ 
where $W'(\lambda_{\alpha})=0$ (simple zeros).  Then $\eta_{ij}=Res
[\phi_i\phi_j/W']$ and $F_{ijk}=\eta_{k\ell}c^{\ell}_{ij}$ with $\phi_1\sim
1$.  This corresponds to standard Whitham theory type WDVV using just the 
$T_n$ times, and can be phrased via differentials $d\Omega_A$ as 
in (\ref{112}).
For the truncated WDVV with only variables $a_i$ involved one writes
$a_i=\oint_{A_i}dS$ with $a^D_i=\oint_{B_i}dS$ and $a_i\sim d\omega_i$ where
the $d\omega_i$ are holomorphic differentials with $\oint_{A_i}d\omega_j=
\delta_{ij}$.  In the present situation one can write the $d\omega_i$ as 
linear combinations of holomorphic differentials (as in (\ref{121})).
Then in \cite{mr} one defines
(note $dw/w\sim dh/h=dP/y$)
\be
F_{ijk}=\frac{\partial^3F}{\partial a_i\partial a_j\partial_k}=
Res_{d\lambda=0}
\left(\frac{d\omega_id\omega_jd\omega_k}{d\lambda(dw/w)}\right)=
\label{123}
\ee
$$=\sum_1^{2g+2}\frac{\hat{\omega}_i(\lambda_{\beta})\hat{\omega}_j
(\lambda_{\beta})\hat{\omega}_k(\lambda_{\beta})}{P'(\lambda_{\beta})/
\hat{y}(\lambda_{\beta})}$$
where $d\omega_i(\lambda)=[\hat{\omega}_i(\lambda_{\beta})+O(\lambda-
\lambda_{\beta})]d\lambda$ and $\hat{y}^2(\lambda_{\beta})=\prod_
{\beta\not=\alpha}(\lambda_{\beta}-\lambda_{\alpha})$.  Note
$dw/w\sim dh/h=dP/y=dy/P$ as in $(\bullet\bullet\bullet)$ and one
asks if $P'd\lambda/y$ is a holomorphic differential; this is assured
via (\ref{121}), so one is truly in the context of \cite{mr}. 
For the metric
one can take
\be
\eta_{ij}(d\omega)=Res_{d\lambda=0}\left(\frac{d\omega_id\omega_jd\omega}
{d\lambda (dw/w)}\right)=\sum\frac{\hat{\omega}_i(\lambda_{\beta})
\hat{\omega}_j(\lambda_{\beta})\hat{\omega}(\lambda_{\beta})}
{P'(\lambda_{\beta})/\hat{y}(\lambda_{\beta})}
\label{124}
\ee
where $d\omega$ is any holomorphic differential.
Then the $c^k_{ij}(d\omega)$ can be obtained via
\be
F_{ijk}=\eta_{k\ell}(d\omega)c^{\ell}_{ij}(d\omega)
\label{125}
\ee
(see below).
These formulas could also be expressed via
\be
F_{ijk}=-Res_{d\,log(w)=0}\left(\frac{d\omega_id\omega_jd\omega_k}
{d\lambda (dw/w)}\right)
\label{126}
\ee
and via $dw/w=dP/y=dE$ the calculation can be taken over
the $q_s$ where $dE(q_s)=0$ (with $d\lambda\sim dQ$).
Thus the formula of (\ref{123}) is compatible with $F_{ABC}$ of (\ref{112})
(and with (\ref{308}))
but the $\eta$ terms (\ref{112}) 
and (\ref{124}) are incompatible since
$\eta_{a_ia_j}=0$ in (\ref{112}) 
($\eta_{ij}=\sum_{q_s}Res[d\omega_id\omega_j/
dE]=0$ since no new residues are introduced at $P_{\pm}$). 
However as indicated in Remark 3.4 the $\eta_{a_ia_j}$ terms in (\ref{500})
can be nonzero so it may be possible to relate them to (\ref{124}) if
e.g. $N(d\Omega_1/d\hat{\Omega}_1)\sim (d\omega/d\lambda)$ can be
achieved (this latter relation is unlikely however
as indicated in Remark 4.1 below - recall $d{\cal E}=(1/N)(dw/w)$). 
In \cite{cb}
we suggested using some sort of combination of prepotentials 
$F_{SW}$ and $F_{Whit}$ to
produce a WDVV theory encompassing both situations 
but this seems excessive.
First consider \cite{kc}.  The formalism is basically
the same but $E$ has the form $E=p^n+u_{n-2}p^{n-2}+\cdots+u_0+O(p^{-1})$
near $P_1\sim P_{+}$.  Unfortunately this is the only point considered
as $\infty$ and the notation does not directly lead to much improvement
over \cite{ko} (and (\ref{WV}), (\ref{112}));
in particular, although $dQ$ is allowed
to have poles at the $P_{\alpha}$ along with $dE$,
no expression for $\eta_{AB}$ is given and our equation (\ref{308}) is 
meant to remedy this situation. 
The full WDVV theory is however
not worked out for this general situation in \cite{kc} so we are left with
\cite{ko} from which the main conclusions (based on (\ref{112})) of interest
here is that $\eta_{a_j,(E,k)}=\delta_{jk}$ and all other $\eta_{A,B}$
involving $a_i,\,a_j$ are zero 
(we have seen however that (\ref{308}) suggests $\eta_{AB}\not= 0$ for
$A,B\sim a_i,\,a_j$ so some intersection of theories may occur).
\\[3mm]\indent
Let us call $F_{SW}$ the prepotential for the truncated theory based on
$a_j$ variables alone as in (\ref{123}) - (\ref{126}) 
(with $a_0\sim\Lambda\sim T_1$ included or not) and $F_W$ the
prepotential arising in the full Whitham theory as in (\ref{WU}),
(\ref{WV}), but for two punctures $P_{\pm}$, so we tentatively
retain the $c_{ABC}$ of (\ref{112}) (or better (\ref{308})).
In \cite{mr,Mw,Ms,Mz} the 
relation of prepotentials is essentially avoided by
working with the $a_j$ variables alone as in (\ref{123}) - (\ref{126});
no attempt is made to connect this with the larger system.  In principle
one might try to extend by using the definition
\be
d\omega_id\omega_j=c^k_{ij}(d\omega)d\omega_kd\omega\,\,mod\left(\frac
{dP}{y^2}d\lambda\right)
\label{135}
\ee leading to (\ref{125}) and setting
\be
\eta_{AB}(d\omega)=Res_{d\lambda=0}\left(\frac{d\Omega_Ad\Omega_Bd\omega}
{d\lambda dE}\right);
\label{136}
\ee
$$d\Omega_Ad\Omega_B=c_{AB}^D(d\omega)d\Omega_Dd\omega\,\,
mod\,\left(\frac{dPd\lambda}{y^2}\right)$$ 
Then
as in \cite{mr,mz}, formally
\be
\eta_{AB}(d\omega)c_{FG}^B(d\omega)=Res_{d\lambda=0}\frac
{d\Omega_A}{d\lambda
dE}\left(d\Omega_Fd\Omega_G+P_{FG}(\lambda)\frac{dPd\lambda}{y^2}\right)=
\label{137}
\ee
$$=F_{AFG}+Res_{d\lambda=0}\left(\frac{P_{FG}(\lambda)d\Omega_A}{y}
\right)=F_{AFG}+\Xi$$ (where $P_{FG}$ is a
polynomial).  Here $\Xi$ is singular at
the zeros of $y$ (branch points) and at poles of
$d\Omega_A$.
In order to have $\Xi=0$ it is thus
required that $Res_{\pm}(P_{FG} d\Omega_A/y)=0$
(which is automatic in \cite{mr} where only
$d\Omega_A$ holomorphic is involved). 
We note that with only $a_j$ variables in (\ref{136}) - (\ref{137}) 
one obtains
(cf. (\ref{135}))
\be
\eta_{k\ell}(d\omega)c^{\ell}_{ij}(d\omega)=Res_{d\lambda=0}\frac
{d\omega_kd\omega_{\ell}d\omega}{d\lambda(dw/w)}c^{\ell}_{ij}(d\omega)=
\label{14}
\ee
$$=Res_{d\lambda=0}\frac{d\omega_k}{d\lambda(dw/w)}\left(d\omega_id\omega_j
-p_{ij}(\lambda)\frac{dPd\lambda}{y^2}\right)=F_{ijk}-Res_{d\lambda=0}\frac
{p_{ij}(\lambda)d\omega_k}{y}$$
Here $d\omega_k=O(1)dz=O(1/\lambda^2)d\lambda$ and $dPd\lambda/y^2=
O(\lambda^g/\lambda^{2g+2})d\lambda^2=O(\lambda^{-g-2})d\lambda^2$ so 
$d\omega_id\omega_j=O(\lambda^{-4})d\lambda^2$ and $p_{ij}O(\lambda^{-g-2})
d\lambda^2$ should be of the same or lower order.  Thus $p_{ij}$ is of 
order $g-2<g+1$ and $Res_{d\lambda=0}[p_{ij}d\omega_k/y]=0$
(cf. \cite{mr} - no residue arises for large $\lambda$).  Suppose we
consider (\ref{136}) - (\ref{137}) in say hyperelliptic parametrization
with differentials $d\Omega^{\pm}_j$
restricted to $d\Omega_1^{\pm}$ plus $d\Omega_0$ and $d\omega_k$.  Then
$d\Omega_1^{\pm}=O(dz/z^2)=O(d\lambda)$ near $P_{\pm}$
and $d\Omega_0=O(dz/z)=
O(d\lambda/\lambda)$.  For $d\Omega_Ad\Omega_B\sim d\Omega_1^{\pm}d\Omega_1^
{\pm}=O(d\lambda)$ one needs $d\Omega_Dd\omega=O(d\lambda)$ in (\ref{136})
and this doesn't work.
Thus it seems that a direct extension of (\ref{135}) to an enlarged context
containing just $d\Omega_1^{\pm}$ and $d\Omega_0$ can not be envisioned;
neither does $\eta_{AB}(d\Omega_1)$ nor $\eta_{AB}(d\Omega_0)$ seem tenable.
\\[3mm]\indent {\bf REMARK 4.1.}$\,\,$
Let us
suggest now that $F_{SW}$ can be $F_{Whit}$ when 
$T_n\to \delta_{n,1}$ and $\alpha_i\to a_i$
(and perhaps $T_1\sim\Lambda\sim a_0$ as in Conclusion 2.1).  
This would take care of $\partial^2F/\partial a_i\partial 
a_j=B_{ij}$ for $F_{SW}$. 
The only difference then arises from
using a different metric $\eta$ with $F_{SW}$ and $F_{Whit}$ and
including more variables in $F_{Whit}$.  One is dealing
with two distinct theories SW and Whitham and for the corresponding
WDVV theories two different metrics are involved.
The Whitham ``geometry" may however interact with the SW
geometry but agreement via $N(d\Omega_1/d\hat{\Omega}_1)=(d\omega/d\lambda)$
(cf. (\ref{500}) and (\ref{124})) seems unlikely.  In this direction
recall from (\ref{309}) that $(d\Omega_1/d\hat{\Omega}_1)=\sum_1^{\infty}
v_p^{\pm}\xi^{p-1}$ near $\infty_{\pm}$ but $dw\sim -\sum_1^{\infty}
\sigma_m^{\pm}\xi^{m-1}d\xi$ as in (\ref{3}) so $\lambda\sim 1/\xi$
implies $d\omega/d\lambda\sim\sum_1^{\infty}\sigma_m^{\pm}\xi^{m+1}$
and one would need $v_1^{\pm}=v_2^{\pm}=0$ which appears unreasonable.

\section{SURVEY OF \cite{Ms,Mz}}
\renewcommand{\theequation}{5.\arabic{equation}}\setcounter{equation}{0}

First in \cite{Mz} one considers the operator ($\psi_z(T)\sim\psi(z,T)$)
\be
{\cal D}_{\mu}(z)\psi=(I\partial_{\mu}-zC_{\mu}(T))\psi(z,T)=0
\label{15}
\ee
where $z$ is a spectral variable, $T=(T^{\mu})$ is some collection of
``times" ($\mu=1,\cdots;\,\,\partial_{\mu}=\partial/\partial T^{\mu}$),
and a variable $T^0$ lurks in the background (see below); mostly here
however the $T^{\mu}\sim T_{\mu}\sim a_{\mu},\,\,1\leq \mu\leq g$.
In (\ref{15}) $C_{\mu}\sim (C_{\mu})^{\alpha}_{\beta}=\eta^{\alpha\gamma}
(F_{\mu})_{\gamma\beta}$ where $F_{\mu\alpha\beta}=
(F_{\mu})_{\alpha\beta}=
\partial^3F/\partial T^{\mu}\partial T^{\alpha}\partial T^{\beta}=
c_{\mu\alpha\beta}$.  The background is of course 
$\phi_i\phi_j=c^k_{ij}\phi_k,
\,\,c^k_{ij}=\eta^{k\ell}c_{ij\ell},\,\,c_{pjk}=\eta_{pi}c^i_{jk}
$, where
$\eta$ is momentarily unspecified.
The associativity condition $(\phi_i\phi_j)\phi_k=\phi_i(\phi_j\phi_k)$
can be written as
\be
\phi_ic_{jk}^{\ell}\phi_{\ell}=c^{\ell}_{jk}\phi_i\phi_{\ell}=
c^{\ell}_{jk}c^m_{i\ell}\phi_m=c^p_{ij}\phi_p\phi_k=c^p_{ij}c^m_{pk}\phi_m
\label{16}
\ee
Since $c^{\ell}_{jk}=c^{\ell}_{kj}$ (via $\eta^{\ell m}c_{mjk}=\eta^
{\ell m}c_{mkj}$) this reduces to $(\bullet\spadesuit\bullet)\,\,
(C_k)^{\ell}_j(C_i)^m_{\ell}=(C_i)^p_j(C_k)^m_p\equiv C_kC_i=C_iC_k$.
Thus the equations (\ref{15}) are referred to as an integrable
(Whitham style) structure behind WDVV, and we remark that this is related
to a well known structure from \cite{cw,dc}, namely
\be
\partial_{\mu}\xi_{\beta}=zc^{\alpha}_{\mu\beta}\xi_{\alpha}
\label{17}
\ee
To see this one can write $\hat{\xi}^i=\eta^{i\alpha}\xi_{\alpha}$ or
$\eta_{jk}\hat{\xi}^i=\delta_{j\alpha}\xi_{\alpha}=\xi_j$ so that 
(\ref{17}) becomes
\be
\partial_{\mu}\xi_{\beta}=z\eta^{\alpha i}c_{i\mu\beta}\xi_{\alpha}=
zc_{i\mu\beta}\hat{\xi}^i\equiv\partial_{\mu}(\eta_{\beta i}\hat{\xi}^i)
=zc_{i\mu\beta}\hat{\xi}^i
\label{18}
\ee
while (\ref{15}) can be written as
\be
\partial_{\mu}\psi^{\alpha}=zc^{\alpha}_{\mu\beta}\psi^{\beta}\equiv
\partial_{\mu}(\eta_{p\alpha}\psi^{\alpha})=zc_{p\mu\beta}\psi^{\beta}
\label{19}
\ee
We state this as
\\[3mm]\indent {\bf PROPOSITION 5.1.}$\,\,$  One can identify 
$\psi^{\alpha}$
in (\ref{15}) or (\ref{19}) with $\hat{\xi}^{\alpha}=\eta^{\alpha p}
\xi_p$ from (\ref{17}).
\\[3mm]\indent {\bf REMARK 5.2.}$\,\,$
Let us recall a few facts from \cite{cw,dc}.  One
deals with equations (\ref{17}) where 
$\partial_1\xi_{\beta}=z\xi_{\beta}$ along with equations
(cf. also \cite{Da})
\be
\partial\psi_{jp}=z\psi_{jp};\,\,\partial_j\psi_{ip}=\gamma_{ij}
\psi_{ip};\,\,
\psi^2_{ip}=\partial_iV^p
\label{34}
\ee
Let us put another index in (\ref{17}) to get
\be
\partial_{\mu}\xi_{\beta}^p=zc^{\alpha}_{\mu\beta}\xi_{\alpha}^p
\label{35}
\ee
and recall that ${\bf (\Delta)}\,\,\xi_i^{\alpha}=\psi_{i1}
\eta^{\alpha p}\psi_
{ip}$ where $\eta_{ij}=\sum_1^N\psi_{mi}\psi_{mj}$ (for $z=0$) and 
$g_{ii}=\psi_{i1}^2$ (for $z=0$).  This latter equation defines an 
Egorov metric when $\gamma_{ij}=\partial_j\sqrt{g_{ii}}/\sqrt{g_{jj}}=
\gamma_{ji}=\partial_i\sqrt{g_{jj}}/\sqrt{g_{ii}}$.
We note here also from \cite{cw} that ${\bf (\Xi)}\,\,\partial_{\alpha}
\partial_{\beta}\tilde{t}=zc^{\epsilon}_{\alpha\beta}\partial_{\epsilon}
\tilde{t}$ where $\xi_{\alpha}=\partial_{\alpha}\tilde{t}$ and a family
$\tilde{t}_{\gamma}$ is involved.  This says in particular (for 
$\partial_{\beta}\tilde{t}_{\gamma}=\xi_{\beta}^{\gamma}$) that
$\partial_{\alpha}\xi_{\beta}^p=\partial_{\beta}\xi_{\alpha}^p=zc^{\epsilon}_
{\alpha\beta}\xi_{\epsilon}^p$ (in agreement with (\ref{17})).  Now we
have identified $\psi^{\alpha}$ in (\ref{15}) with $\hat{\xi}^{\alpha}=
\eta^{\alpha s}\xi_s$ and one asks for possible relations between these
$\psi^{\alpha}$ and the $\psi_{jp}$ of ${\bf (\Delta)}$.  
It would seem that an 
index $p$ could be attached here to $\xi_s$ so we define 
${\bf (\Upsilon)}\,\,
\hat{\xi}_p^{\alpha}=\eta^{\alpha s}\xi_s^p$.  Then denoting the $\psi$
of (\ref{15}) by $\tilde{\psi}$ one can conclude from ${\bf (\Delta)}$,
setting $\xi_i^{\alpha}=\eta_{im}\hat{\xi}^m_{\alpha}$,
\be
\psi_{i1}\eta^{\alpha p}\psi_{ip}=\psi_{i1}\psi^{\alpha}_i=\xi_i^{\alpha}
=\eta_{im}\hat{\xi}^m_{\alpha}=\eta_{im}\tilde{\psi}_{\alpha}^m\equiv
\sqrt{g_{ii}}\psi^{\alpha}_i=\eta_{im}\tilde{\psi}^m_{\alpha}
\label{36}
\ee
where one posits $N$ independent solutions $\tilde{\psi}_m$ of (\ref{15}).
\\[3mm]\indent
We hope to develop the meaning of this later 
but continue now to sketch results in \cite{Ms,Mz}.
The WDVV equations for truncated $F=F_{SW}$ depend only on $a_j\,\,
(1\leq j\leq g)$ and they are involved in a more general formulation
\be
F_iF_k^{-1}F_j=F_jF_k^{-1}F_i;\,\,C^j_j=F_k^{-1}F_j;\,\,[C_i^k,C_j^k]=0
\label{20}
\ee
(note the first equation corresponds to $F_kC^k_iC^k_j=F_kC^k_jC^k_i$).  
If (\ref{20}) is true for one $k$ (say $k=0$) with $[c^0_i,C^0_j]=0$ and
all the $F_k$ are nondegenerate so that $F_k^{-1}$ exists then write
$F_j=F_0C^0_j$ with $F_iF_k^{-1}F_j=F_0C^0_i(C^0_k)^{-1}F_0^{-1}F_0C^0_j=
F_0(C^0_i(C^0_k)^{-1}C^0_j)$ and this is the same as $F_jF_k^{-1}F_i=
F_0(C^0_j(C^0_k)^{-1}C^0_j)$ (attributed to A. Rosly).  Indeed from
$AB=BA$ and $CB=BC$ one has $ABC=BAB^{-1}BC=BAC$ with $CBA=BCB^{-1}B=BCA$.
Now generally (\ref{20}) may not include a constant, moduli independent,
$F_{\lambda}$ and hence may not imply (\ref{15}); therefore in \cite{Mz}
(based apparently on \cite{dc}) one takes instead the equations
\be
(\partial_{\mu}-C^{\lambda}_{mu}\partial_{\lambda})\psi=0\equiv
(F_{\lambda}\partial_{\mu}-F_{\mu}\partial_{\lambda})\psi=0
\label{21}
\ee
for all $\lambda,\,\mu$ (note here $C^{\lambda}_{mu}=F_{\lambda}^{-1}
F_{\mu}$ for the equivalence).  One checks that the operators 
$\partial_{\mu}-C^{\lambda}_{\mu}\partial_{\lambda}$ with different 
$\lambda,\,\,\mu$ commute.  Further the equations are consistent
$(\psi$ can be chosen independently of $\lambda$) since
\be
F_{\mu}\partial_{\nu}-F_{\nu}\partial_{\mu}=F_{\mu}(\partial_{\nu}
-C^{\lambda}_{\nu}\partial_{\lambda})-F_{\nu}(\partial_{\mu}
-C^{\lambda}_{\mu}\partial_{\lambda})
\label{22}
\ee
(note $F_{\mu}C^{\lambda}_{nu}-F_{\nu}C^{\lambda}_{mu}=F_{\mu}F_{\lambda}^
{-1}F_{\nu}-F_{\nu}F_{\lambda}^{-1}F_{\mu}$).  Then in order to arrive
at (\ref{15}) from the generic system (\ref{21}) 
take $\psi=exp(zT^0)\psi_z$ which is a self
consistent ansatz when the $C_{\mu}^0$ are independent of $T^0$.
\\[3mm]\indent
Next go to \cite{Ms} where one takes
\be
F_iG^{-1}F_j=F_jG^{-1}G_i;\,\,G=\sum_1^r\eta^k(T)F_k
\label{23}
\ee
where the $F_i$ are $r\times r$ matrices with $(F_i)_{jk}=F_{ijk}$ as usual
and the $\eta^k(T)$ are (row) vectors.  Via \cite{By} one can also
use ($T^i\sim T_i$)
\be
{\cal F}(T^0,T^1,\cdots,T^r)=T_0^2F(T^i/T^0)
\label{24}
\ee
and writing ${\cal G}=\sum_0^r\eta^K{\cal F}_K$ with $(r+1)\times
(r+1)$ matrices ${\cal F}_K$ and $\hat{{\cal G}}^{-1}=(det{\cal G})
{\cal G}^{-1}$ the WDVV equations can be rewritten as
\be
{\cal F}_I\hat{{\cal G}}^{-1}{\cal F}_J={\cal F}_J\hat{{\cal G}}^{-1}
{\cal F}_I;\,\,I,J=(0,1,\cdots,r)
\label{25}
\ee
Further $T^0{\cal F}_{0ij}=-{\cal F}_{ijk}T^k,\,\,T^0{\cal F}_{00i}=
{\cal F}_{ik\ell}T^kT^{\ell},\,\,T^0{\cal F}_{000}=-{\cal F}_{k\ell m}
T^kT^{\ell}T^m$, etc.  As in (\ref{21}) the WDVV equations imply
the consistency of
\be
\left(F_{ijk}\frac{\partial}{\partial T^{\ell}}-F_{ij\ell}\frac{\partial}
{\partial T^k}\right)\psi^j(T)=0
\label{26}
\ee
and contracting with $\eta^{\ell}(T)$ this can be rewritten as
\be
\frac{\partial \psi^i}{\partial T^k}=c^i_{jk}D\psi^j;\,\,C_k=G^{-1}F_k;\,\,
G=\eta^{\ell}F_{\ell};\,\,D=\eta^{\ell}\partial_{\ell}
\label{27}
\ee
The equations (\ref{23}) can be rewritten as $[C_i,C_j]=0$ and they
are invariant under linear changes of $T^k$ with $F$ fixed.  There are
also nonlinear transformations preserving the WDVV structure but they
change the prepotential.  It is shown in \cite{Ms} that such transformations
are naturally induced by solutions of (\ref{26}) via ${\bf (W)}\,\,
T^i\to\tilde{T}^i=\psi^i(T);\,\,F(T)\to\tilde{F}(\tilde{T})$ and the
period matrix remains intact, i.e. ${\bf (X)}\,\,F_{ij}=\partial^2/
\partial T^i\partial T^j=\partial^2\tilde{F}/\partial\tilde{T}^i
\partial\tilde{T}^j$.  If there is a distinguished time variable
$T^r$ such that all $F_{ijk}$ are independent of $T^r$ (i.e. $\partial_r
F_{ijk}=0=\partial_iF_{rjk}$), then set $\tilde{\psi}^k_z(T^1,\cdots,
T^{r-1})=\int\psi_z^k(T^1,\cdots,T^{r-1},T^r)exp(zT^r)dT^r$ (Fourier
transform) to obtain
\be
\frac{\partial\tilde{\psi}^i_z}{\partial T^j}=zc^i_{jk}\tilde{\psi}^k_z
\label{28}
\ee
(take $k=r$ in (\ref{26}) and $\ell=j$ to get $(F_r\partial_j-F_j
z)\tilde{\psi}=0$ with $F_r^{-1}F_j\sim C_j$).  In this case 
$T^i\to\tilde{T}^i_z=\tilde{\psi}^i_z(T)$.
\indent
Next one looks at infinitesimal variations of WDVV (\ref{23}) which preserve
their shape.  One writes ${\bf (Y)}\,\,\tilde{F}(T)=F(T)+\epsilon f(T)$ and
$T^i=\tilde{T}^i+\epsilon\xi^i(T)$.  This leads to
\be
\tilde{F}_{ijk}=F_{ijk}+\epsilon\left(\left\{f+
\partial_{\ell}F\xi^{\ell}\right\}_{ijk}
-F_{ijk\ell}\xi^{\ell}\right)
\label{29}
\ee
and the right side corresponds to elements of $F_j(1+\epsilon A_j)$
where 
\be
F_{ijn}A^n_k\equiv \left\{f+\partial_{\ell}F\xi^{\ell}\right\}_{ijk}-
F_{ijk\ell}\xi^{\ell}
\label{30}
\ee
Then the form of WDVV (\ref{23}) is preserved by ${\bf (Y)}$ provided
${\bf (Z)}\,\, F_i(A_i-A_j)F_i^{-1}=F_k(A_k-A_j)F_k^{-1}$.  Any
constant matrices $A_i=A$ will work in which case (subscripts
on $\xi$ denote partial derivatives)
\be
F_{ijn}A^n_k=\{f+\partial_{\ell}F\xi^{\ell}\}_{ijk}-F_{ijk\ell}\xi^{\ell}=
f_{ijk}+(F_{i\ell}\xi^{\ell}_j+F_{\ell j}\xi^{\ell}_i+\partial_
{\ell}F\xi^{\ell}_{ij})_k+F_{ij\ell}\xi^{\ell}_k
\label{31}
\ee
The last term on the right is of the desired form so one demands
that ${\bf (\Gamma)}\,\,F_{i\ell}\xi^{\ell}_j+F_{\ell j}\xi^{\ell}_i
+\partial_{\ell}F\xi^{\ell}_{ij}=-f_{ij}$.  This is however exactly
the infinitesimal variation of the period matrix so the invariance
of the period matrix is the condition for infinitesimal WDVV
covariance.

\section{KERNELS}
\renewcommand{\theequation}{6.\arabic{equation}}\setcounter{equation}{0}

The results of \cite{ch} show that dKdP is characterized by the kernel
in (\ref{C}).  Thus dKdP can be defined via ${\bf (\Theta)}\,\,
{\cal B}_n(P)=\lambda_{+}^n=\sum_0^nb_{nm}P^m=\partial_nS$
which implies that $\partial_nP=\partial{\cal B}_n$, while the approach
of \cite{ta} characterizes dKdP via the dispersionless differential Fay
identity (cf. also \cite{ci}) which is shown in \cite{ch} to be
equivalent to the kernel (\ref{C}).  The version of $K$ appropriate
to a RS (we take one point at infinity or one puncture for convenience)
is then obtained as in \cite{cf} via ${\cal B}_n\sim\Omega_n\sim
\int^{\gamma}d\Omega_n$ with $\int^{\gamma}d\Omega_1=\Omega_1\sim p$
corresponding to $P=S_X$ (here $\gamma\in\Sigma_g$ is a generic point).
Thus in (\ref{C}) $\partial_PQ_n\sim(1/n)(\partial\Omega_n/
\partial\Omega_1)$ and we can write
\be
\frac{\partial Q_n}{\partial P}\sim\frac{1}{n}\frac{\partial\Omega_n/
\partial\gamma}{\partial\Omega_1/\partial\gamma}=\frac{1}{n}
\frac{d\Omega_n}{d\Omega_1}
\label{400}
\ee
so for $\lambda,\,\mu$ near $\infty$
\be
K(\mu,\lambda)\sim{\cal K}(\mu,\lambda)=\sum_1^{\infty}\frac{1}{j}
\frac{d\Omega_j}{d\Omega_1}(\lambda)\mu^{-j}=\frac{1}{d\Omega_1(\lambda)}
\sum_1^{\infty}d\Omega_j(\lambda)\frac{\mu^{-j}}{j}
\label{401}
\ee
Thus ${\cal K}$ is a generating function for differentials $d\Omega_j$ and to 
compare this with $W$ of (\ref{71}) one would write first for $\xi\sim
1/\lambda$ and $\zeta\sim 1/\mu$ with $\lambda,\,\mu\to\infty$ the
identity $d\Omega_j(\lambda)\sim d\Omega_j(\xi)$ for $d\Omega_j(\xi)\sim
d\Omega^{+}_j(\xi)$ as in (\ref{130}) so $d\Omega_j(\xi)\sim-jd\dot
{\Omega}_j(\xi)$ where $d\dot{\Omega}_j$ is the differential used in
\cite{gf}.  Hence ${\cal K}(\mu,\lambda)\sim{\cal K}(\zeta,\xi)$ where
\be
{\cal K}(\zeta,\xi)\sim -\frac{1}{d\Omega_1(\xi)}\sum_1^{\infty}
d\dot{\Omega}_j(\xi)\zeta^j
\label{402}
\ee
leading to
\be
d_{\zeta}{\cal K}(\zeta,\xi)=-\frac{1}{d\Omega_1(\xi)}\sum_1^{\infty}
j\zeta^{j-1}d\dot{\Omega}_j(\xi)d\zeta=
-\frac{W(\xi,\zeta)}{d\Omega_1(\xi)}
\label{403}
\ee
Following \cite{cf} we can write further
\be
d\Omega_1(\xi){\cal K}(\zeta,\xi)=\frac{d\xi}{\xi}-d_{\xi}log\,E(\xi,\zeta)
\label{404}
\ee
where ${\bf (\Sigma)}\,\,d_{\xi}log\,E(\xi,\zeta)={\cal Z}_{\zeta}(\xi)$
is a local Cauchy kernel on $\Sigma_g$ and formally this becomes
($p\sim\Omega_1$)
\be
\frac{dp(\xi)}{p(\zeta)-p(\xi)}=\frac{d\xi}{\xi}-{\cal Z}_{\zeta}
(\xi);
\label{405}
\ee
$${\cal Z}_{\zeta}(\xi)\sim\left[\frac{1}{\xi-\zeta}+\sum_1^{\infty}
\frac{q_{ms}}{s}\xi^{m-1}\zeta^s\right]d\xi$$
($q_{ms}$ as in (\ref{130}) for $d\Omega_j\sim d\Omega_j^{+}$).
In any event we know that the $d\Omega_n$ or $d\dot{\Omega}_n$ are
determined by the Szeg\"o kernel on $\Sigma_g$ and (working in the one
puncture situation for convenience) we use (\ref{130}) for 
$d\Omega_n\sim d\Omega_n^{+}$ (normalized via $\oint_{A_j}d\Omega_n=0$)
with (\ref{3}) for $d\omega_j$.  Then (cf. \cite{cc,na,sc}) we recall that
${\bf (\Phi)}\,\,\oint_{B_j}d\Omega_n=\Omega_{nj}=-2\pi iRes\,z^{-n}
d\omega_j=2\pi i\sigma_{jn}$ which means that the Szeg\"o kernel also
determines the holomorphic differentials $d\omega_j$.  Hence as in
Conclusion 2.1 one has
\\[3mm]\indent {\bf CONCLUSION 6.1.}$\,\,$  For Whitham theories based
on times $T_n$ and $\alpha_j$ alone (as in (\ref{73}) - (\ref{74}))
the Szeg\"o kernel on a given $\Sigma_g$ (or ${\cal K}$ of 
(\ref{401}) - (\ref{402})) determines the Whitham prepotential $F_{Whit}$
and action differential $dS$.  Conversely from $d\Omega_n=\partial dS/
\partial T_n$ etc. one can recover the Szeg\"o kernel in terms of the
full Whitham $dS$, as well as the $d\omega_j$ and $\alpha_j$, and this 
represents some kind of weak determination of $\Sigma_g$ (the canonical
homology basis and the complex structure are unspecified for example
and the ramification is undetermined).  Given a Toda type
hyperelliptic curve $\Sigma_g$ the determination of moduli $u_k$ or 
$h_j$ should suffice for specification.  In terms of prepotential
$F_{Whit}$ it is evidently determined by the RS as indicated before.
Conversely $F_{Whit}$ determines the period matrix $B_{ij}$ via
(\ref{300}) and in fact the expansion (\ref{78}) can be written out as in
\cite{na} in terms of the $q_{ij},\,\,\sigma_{ij}$, and $B_{ij}$ for
normalized differentials $d\Omega_n^{\pm}$ and $d\omega_j$, thus
giving the same kind of information as just indicated for $dS$.

\end{document}